\documentclass[english,11pt,a4paper]{article}
\usepackage[T1]{fontenc}
\usepackage[latin9]{inputenc}
\usepackage{mathrsfs}
\usepackage{amsmath}
\usepackage{amsthm}
\usepackage{amssymb}
\usepackage{graphicx}

\makeatletter
\numberwithin{equation}{section}
\numberwithin{figure}{section}

\usepackage{jheppub}

\makeatother

\usepackage{babel}
\begin{document}
\preprint{ULB-TH/22-04}
\title{Neutrino forces and the Sommerfeld enhancement}

\abstract{The Sommerfeld enhancement plays an important role in dark matter (DM) physics, and can significantly enhance the annihilation cross section of non-relativistic DM particles. In this paper, we study the effect of neutrino forces, which are generated by the exchange of a pair of light neutrinos,  on the Sommerfeld enhancement. We demonstrate that in certain cases, a neutrino force can cause a significant correction to the Sommerfeld enhancement. Models that can realise DM-neutrino interactions and sizeable Sommerfeld enhancement are also briefly discussed, together with the impacts on DM phenomenology of neutrino forces.}

\author[a]{Rupert Coy,} 
\emailAdd{Rupert.Coy@ulb.be}
\author[b,c]{Xun-Jie Xu,}
\emailAdd{xuxj@ihep.ac.cn}
\author[b,c]{Bingrong Yu} 
\emailAdd{yubr@ihep.ac.cn}
\affiliation[a]{Service de Physique Th\'{e}orique, Universit\'{e} Libre de Bruxelles, Boulevard du Triomphe, CP225, 1050 Brussels, Belgium}
\affiliation[b]{Institute of High Energy Physics, Chinese Academy of Sciences, Beijing 100049, China}
\affiliation[c]{School of Physical Sciences, University of Chinese Academy of Sciences, Beijing 100049, China}
\maketitle

\section{Introduction}
The Sommerfeld enhancement is a non-relativistic quantum effect that can significantly change the annihilation cross section of two slow-moving particles~\cite{Sommerfeld}. When the two particles interact through the exchange of a 
light mediator, 
the plane-wave approximation of the incident particles is violated and there can be a significant enhancement (or suppression) of the annihilation cross section due to the attractive (or repulsive) force which is mediated. 
This is given by
\begin{eqnarray}
\sigma=S\sigma_{\rm free}\;,
\end{eqnarray}
where $\sigma$ and $\sigma_{\rm free}$ are the annihilation cross sections with and without the interaction, respectively. The dimensionless factor $S$ is known as the Sommerfeld enhancement factor that 
quantifies the enhancement (or suppression).

The importance of the Sommerfeld enhancement in the dark matter (DM) annihilation was first realised in Refs.~\cite{Hisano:2003ec,Hisano:2004ds,Hisano:2005ec,Hisano:2006nn,Cirelli:2007xd,March-Russell:2008lng}. 
Originally motivated 
by the observation of cosmic positron excesses~\cite{PAMELA:2008gwm,Chang:2008aa,Fermi-LAT:2009yfs}, there have been a number of works focusing on the effects of the enhancement on DM annihilation and indirect detection~\cite{Cirelli:2008jk,Cirelli:2008pk,Arkani-Hamed:2008hhe,Pospelov:2008jd,Fox:2008kb,Lattanzi:2008qa,Pieri:2009zi,Bovy:2009zs,Yuan:2009bb,Slatyer:2009vg,Feng:2009hw,Feng:2010zp,Cholis:2010px}. 
Many of these studies use the Sommerfeld enhancement generated by Yukawa potentials to enhance indirect detection signals.
On the other hand, if the DM abundance is determined by the freeze-out mechanism in the early universe, the Sommerfeld enhancement can also modify the annihilation rate of DM, 
and thus 
alter its abundance~\cite{Hisano:2006nn,Cirelli:2007xd,March-Russell:2008lng,Zavala:2009mi,Hannestad:2010zt,Iminniyaz:2010hy,Iminniyaz:2011pva,Hryczuk:2011tq}.

Due to its important applications to DM studies, the Sommerfeld enhancement has been extensively investigated in various aspects. 
The Sommerfeld enhancement factors for partial waves of higher orders than the $s$-wave were calculated in Refs.~\cite{Iengo:2009xf,Iengo:2009ni,Cassel:2009wt}, and turn out to be relevant in the non-perturbative region or under some special mechanism~\cite{Das:2016ced}.
The force mediators that lead to enhancements of the DM annihilation can be, 
e.g., pseudo-scalar Goldstone bosons~\cite{Bedaque:2009ri}, unparticles~\cite{Chen:2009ch} and multiple mediators~\cite{McDonald:2012nc,Zhang:2013qza}. 
Additionally, the annihilation rates of the heavy electroweak triplets which can explain neutrino masses in the type-II and type-III seesaw mechanisms may also be enhanced by the Sommerfeld corrections, thus will significantly reduce the baryon asymmetries they generate via leptogenesis~\cite{Strumia:2008cf}. 

Generally speaking, in the presence of light particles mediating long-range forces,  the Sommerfeld enhancement is expected. 
Neutrinos, as one of the lightest species in the Standard Model, 
might also serve as a light mediator for the Sommerfeld enhancement.  It is well known that the exchange of a pair of 
light neutrinos between two particles leads to a long-range force~\cite{Feinberg:1968zz,Feinberg:1989ps, Hsu:1992tg}. 
The effective potential of such a force is proportional to $1/r^5$ if we neglect neutrino masses and assume a contact interaction\footnote{When the distance $r$ exceeds the inverse of neutrino masses, the potential becomes exponential suppressed~\cite{Grifols:1996fk};  when $r$ 
 is too small to maintain the contact vertex, the  $1/r^5$ form also changes to other forms~\cite{Xu:2021daf}.}.
Neutrino forces might have important cosmological and astrophysical effects. Early studies include long-range forces from the cosmic neutrino background~\cite{Hartle:1970ug,Horowitz:1993kw,Ferrer:1998ju} 
and the possibility of neutron stars being affected by neutrino forces~\cite{Fischbach:1996qf,Smirnov:1996vj,Abada:1996nx,Kachelriess:1997cr,Kiers:1997ty,Abada:1998ti,Arafune:1998ft}. 
More recently, Ref.~\cite{Orlofsky:2021mmy} showed 
that the neutrino force generated from a DM-neutrino interaction could be strong enough to impact small-scale structure formation in the early universe.

So far, the possibility that the Sommerfeld enhancement could be caused by neutrino forces has not been discussed in the literature. 
In the present paper, we investigate the effects of neutrino forces on the Sommerfeld enhancement. 
This may be significant in the presence of DM-neutrino interactions, and may therefore affect the DM thermal evolution and DM indirect detection today.  
DM-neutrino interactions are worthy of consideration for several reasons. 
They are present in many models of DM, such as sterile neutrino DM, neutrino portal models or other scenarios which link the dual problems of neutrino masses and DM. 
On the observational side, DM-neutrino couplings are generically less constrained than DM couplings with other SM particles while their cosmological phenomenology is rich~\cite{Wilkinson:2014ksa,Bertoni:2014mva,Olivares-DelCampo:2017feq,Berlin:2017ftj,Stadler:2019dii,Hufnagel:2021pso} 
and 
 provides an additional way to test these, albeit indirectly. 
Moreover, there could be significant implications for structure formation, indeed some cosmological data seems to favour such interactions~\cite{Hooper:2021rjc}.

The remainder of this paper is organised as follows. 
In Sec.~\ref{sec:SE}, we briefly review the Sommerfeld enhancement and the formalism to calculate it for a general potential. 
We then calculate the Sommerfeld enhancement factor in the specific case of a neutrino force in Sec.~\ref{sec:NuForce}. 
As we show, a straightforward calculation using a potential more singular than $1/r^2$ in the Schr\"odinger equation is invalid, thus one cannot use the contact $1/r^5$ neutrino force at all scales. Rather, one should open up the contact vertex and consider the short-range behaviour of neutrino forces in the calculation of the Sommerfeld enhancement. 
We consider two possible 
UV completions of the contact interaction and calculate the corresponding Sommerfeld enhancement factor in Secs.~\ref{subsec:t-channel} and \ref{subsec:s-channel}, where the short-range behaviour of the neutrino force is $1/r$ and $1/r^2$, respectively~\cite{Xu:2021daf}. 
Then in Sec.~\ref{sec:DM pheno}, we discuss how to build models that can realise the DM-neutrino interaction and generate sizeable Sommerfeld enhancements, as well as the impacts on DM phenomenology of neutrino forces. 
We summarise our main results in Sec.~\ref{sec:summary}. Finally, some mathematical aspects about the short-range behaviour of the radial wave function in a general inverse-power potential are discussed in Appendix~\ref{sec:Small-r}.

\section{Sommerfeld enhancement}
\label{sec:SE}

The Sommerfeld enhancement occurs in processes involving long-range
attractive forces between two slow-moving particles. In quantum field
theory, it is a non-perturbative effect which can  be computed by
solving the Bethe-Salpeter equation~\cite{Iengo:2009ni,Cassel:2009wt}. Since
the enhancement itself is only related to soft scattering of slow-moving
particles, it can be computed in nonrelativistic quantum mechanics.
In this section, we briefly review the quantum mechanics approach
to the Sommerfeld enhancement~\cite{Arkani-Hamed:2008hhe}. 

Consider the collision of two particles affected by a long-range
attractive potential $V$. Denote the wave function of the two particles with coordinates $\mathbf{r}_{1}$ and $\mathbf{r}_{2}$ by $\Psi(\mathbf{r}_{1},\ \mathbf{r}_{2})$. 
The wave function
can be written as a product of two parts that depend on $\mathbf{r}\equiv\mathbf{r}_{1}-\mathbf{r}_{2}$
and $\mathbf{r}_{1}+\mathbf{r}_{2}$ respectively. The part depending
on $\mathbf{r}_{1}+\mathbf{r}_{2}$ is not important since it is merely
a plane-wave solution describing the motion of the center-of-mass.
The $\mathbf{r}$-dependent part, denoted by $\psi(\mathbf{r})$,
is however affected by the potential. Specifically, it is determined by
the following  Schr\"odinger equation:
\begin{equation}
\left[-\frac{\nabla^{2}}{2\mu}+V(r)\right]\psi(\mathbf{r})=E\psi(\mathbf{r})\thinspace,\label{eq:-5}
\end{equation}
where $\mu=m_{1}m_{2}/(m_{1}+m_{2})$ is the reduced mass of the two
particles, whose individual masses are $m_{1}$ and $m_{2}$. 
The boundary condition is given by $\psi\to e^{ikz}+f(\theta)e^{ikr}/r$
as $r\to\infty$, where $e^{ikz}$ corresponds to an incoming plane
wave along the $z$-axis and $f(\theta)e^{ikr}/r$ to an outgoing
spherical wave. 

The Sommerfeld enhancement factor is determined by the ratio\footnote{{For a more rigorous treatment of the short-range behaviour of the wave function in the calculation of the Sommerfeld enhancement, see Ref.~\cite{Blum:2016nrz}.}}:
\begin{equation}
S=\left|\frac{\psi(0)}{\psi_{{\rm free}}(0)}\right|^{2},\label{eq:-9}
\end{equation}
where $\psi$ and $\psi_{{\rm free}}=e^{ikz}$ denote solutions of
Eq.~\eqref{eq:-5} with and without $V$, respectively. 

Since the solutions are symmetric around the $z$-axis, we expand
it in Legendre polynomials,
\begin{equation}
\psi(\mathbf{r})=\sum_{l=0}^{\infty}P_{l}(\cos\theta)\frac{u_{l}(r)}{r}\thinspace.\label{eq:-8}
\end{equation}
Since $e^{ikz}=\sum_{l=0}^{\infty}P_{l}(\cos\theta)i^{l}j_{l}(kr)$,
 the expansion of $\psi_{{\rm free}}$ simply gives $u_{{\rm free},\thinspace l}(r)=r i^{l} j_{l}(kr)$.
In particular, for $l=0$, we have
\begin{equation}
u_{{\rm free},\thinspace0}(r)=rj_{0}(kr)=\frac{1}{k}\sin(kr)\thinspace.\label{eq:-14}
\end{equation}
It has been shown in Ref.~\cite{Arkani-Hamed:2008hhe} that contributions of $l\geqslant 1$
to the Sommerfeld enhancement factor vanish if the potential does
not blow up faster than $1/r$ near the origin\footnote{{In fact, if the potential blows up faster than $1/r$
but slower than $1/r^{2}$, or if $\lim_{r\to0}r^{2}|V(r)|$ is a
small (smaller than a certain critical  value) finite value, this
conclusion still holds. For further discussions, see Appendix~\ref{sec:Small-r}.}}. Hence, we will only focus on the $l=0$ mode. For simplicity, in what
follows,  we  denote $u_{0}(r)$ by $u(r)$, and $u_{{\rm free},\thinspace0}(r)$
by $u_{{\rm free}}(r)$. 

Applying the Legendre expansion in Eq.~\eqref{eq:-8} to Eq.~\eqref{eq:-5},
we obtain the radial Schr\"odinger equation for $\chi(r)$:
\begin{equation}
-\frac{1}{2\mu}u''(r)+V(r)u(r)=Eu(r)\ .\label{eq:-7-1}
\end{equation}
The Sommerfeld enhancement factor is then determined by $u$, 
\begin{equation}
S=\lim_{r\to0}\left|\frac{u(r)/r}{u_{{\rm free}}(r)/r}\right|^{2}=\left|u'(0)\right|^{2}.\label{eq:-10}
\end{equation}
Note that the amplitude of $u(r)$ at $r\rightarrow\infty$ is
fixed by $u_{{\rm free}}(r)$, i.e.~$u(r)$  should be identical
to $\sin(kr)/k$ up to a phase shift.  Therefore,  one may need to
 solve Eq.~\eqref{eq:-7-1} with the initial condition $u(0)=0$
and $u'(0)=C$ where $C$ is  determined by matching the amplitude
of $u(r)$ to $u_{{\rm free}}(r)$ at $r\rightarrow\infty$.
Since Eq.~\eqref{eq:-7-1} is linear with respect to $u$, changing
$u'(0)$ by a factor of $C$ corresponding to multiplying the entire
$u(r)$ function by the same factor. Hence one can start with the
initial value $u'(0)=1$, 
solve Eq.~\eqref{eq:-7-1} to get the
amplitude, $A_{u}$,  and then compute the Sommerfeld enhancement
factor by
\begin{equation}
S=\left|\frac{1}{kA_{u}}\right|^{2}.\label{eq:-13}
\end{equation}

\begin{figure}
\centering

\includegraphics[width=0.99\textwidth]{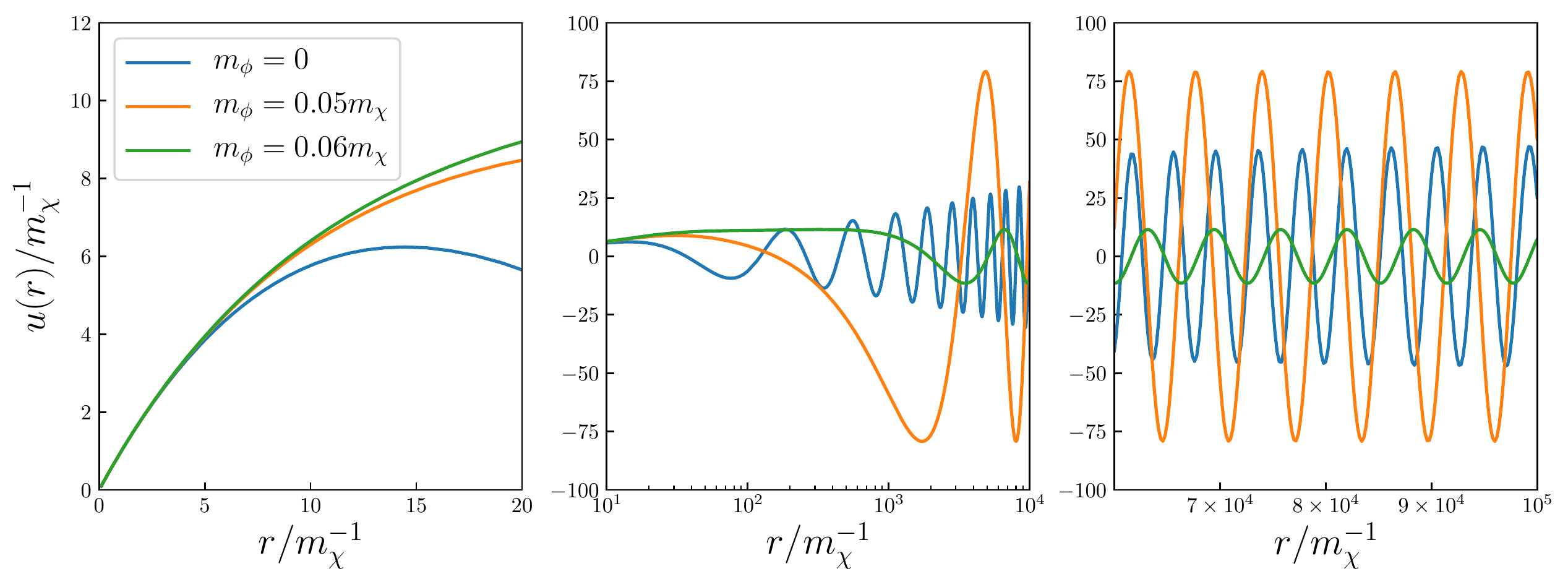}
\caption{Examples of the  solutions of the radial Schr\"odinger equation
for  Yukawa potentials. \label{fig:sol} 
}
\end{figure}

Let us apply the above procedure to the well-studied case of Yukawa
potentials. Consider a Yukawa interaction ${\cal L}\supset - y_{\chi}\phi\overline{\chi}\chi$,
where $\phi$ is a light scalar with mass $m_{\phi}$ and $\chi$ a heavy fermion with mass $m_\chi$. 
The interaction induces the potential, 
\begin{equation}
V_{\phi}=-\frac{\alpha_\chi}{r}e^{-m_{\phi}r}\thinspace,\label{eq:-12}
\end{equation}
with $\alpha_\chi \equiv y_\chi^{2}/(4\pi)$.

Substituting Eq.~\eqref{eq:-12} into Eq.~\eqref{eq:-7-1}, we numerically
solve it with $k/m_\chi =10^{-3}$, $\alpha_\chi =0.1$, and $m_{\phi}=\{0,\ 0.05,\ 0.06\}m_\chi$,
as shown in Fig.~\ref{fig:sol}. 
The choice of $k/m_\chi = v = 10^{-3}$ is motivated by the fact that DM velocities are $v_{\rm DM} \sim 10^{-3}$ today. 
As one can see, starting from the
origin, $u(r)$ initially increases linearly with $r$, then oscillates with an increasing amplitude, and eventually  behaves as a free particle with a constant amplitude. 
According to Eq.~\eqref{eq:-13}, the smaller the final amplitude, the larger the value of $S$ we obtain.  Substituting the obtained amplitudes into
Eq.~\eqref{eq:-13}, we find $S= 445.3$, 159.5, and 7541.2 for the blue, orange,  and green curves, respectively.

\section{Effects of neutrino forces on Sommerfeld enhancements \label{sec:NuForce}}

\begin{figure}
	\centering
	
	\includegraphics[width=0.75\textwidth]{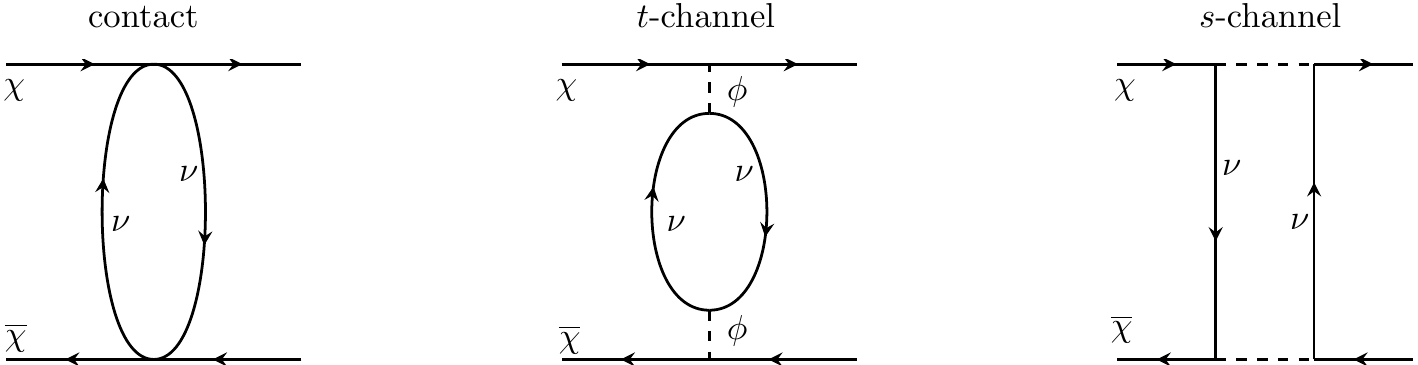}\caption{Feynman diagrams of neutrino forces arising from contact (left), $t$-channel
	(middle), and $s$-channel (right) interactions. \label{fig:feyn} 
	}
\end{figure}

Like other light particles, neutrinos 
may mediate long-range forces as well. Since neutrinos are fermions, the exchange of a pair of neutrinos is required to generate a force,
as shown in Fig.~\ref{fig:feyn}. For contact interactions (left
panel of Fig.~\ref{fig:feyn}), the induced potential is proportional
to $1/r^{5}$, provided that neutrino masses are negligible. The $1/r^{5}$
form remains valid as long as the external fermions are non-relativistic
and the momentum transfer between them, $q\sim r^{-1}$, is smaller
than the energy scale of the contact interaction. At smaller distances
(correspondingly higher momentum transfer), the $1/r^{5}$ form will 
be modified to $1/r^{2}$ or $1/r$, depending on possible ways of
opening the contact interaction~\cite{Xu:2021daf}. In this section,
we investigate possible effects of neutrino forces  on the Sommerfeld enhancement. 
We will first consider the case of contact interactions, for which it is necessary to impose a cut-off at some small distance
scale. Then we consider possible modifications of the potential in
the $t$- and $s$-channel cases, as shown in the middle and right
panels of Fig.~\ref{fig:feyn}.  We will show that the Sommerfeld enhancement in the contact interaction scenario is in general quite weak, unless the contact interaction strength is greater than $1/m_{\chi}^{2}$, above which the UV completion becomes important. It turns out that for the $s$-channel UV completion, the Sommerfeld enhancement remains weak, while for the $t$-channel case, the enhancement can be quite significant and the contribution of neutrino forces is more than merely a loop correction to the tree-level mediator. 

\subsection{Contact interactions\label{subsec:Contact}}

For illustration, we consider the following scalar type of contact
interactions:

\begin{equation}
{\cal L}_{{\rm int}}\supset G_{S}\overline{\nu}\nu\overline{\chi}\chi\thinspace,\label{eq:L-con}
\end{equation}
where $G_{S}$ is the effective coupling strength, $\nu$ denotes
a neutrino and $\chi$ is a fermion, which we may imagine is dark matter. 
One may consider other types of interactions such as pseudoscalar, vector, axial-vector, or tensor. In such generalizations, neutrino forces are either suppressed (e.g.~pseudoscalar) or of the same order of magnitude (e.g.~vector) as the scalar case, depending on whether the non-relativistic limit of the $\chi$ bilinear is velocity-suppressed or not. 

Eq.~\eqref{eq:L-con}
leads to, via the first diagram in Fig.~\ref{fig:feyn}, the following
attractive potential~\cite{Feinberg:1968zz,Xu:2021daf}:
\begin{equation}
V_{c}=-\frac{3G_{S}^{2}}{8\pi^{3}r^{5}}\thinspace.\label{eq:V-con}
\end{equation}

Note that the power of $r$ in the denominator is greater than two.
For an attractive potential $V\propto1/r^{m}$ with $m>2$,  a straightforward
calculation using the potential in the Schr\"odinger equation  would
be invalid. This has previously been addressed in a series of studies
on the theory of singular potentials---see Ref.~\cite{Frank:1971xx}
for a review. That the critical value of the power is two can be understood
using Landau and Lifshitz's argument~\cite{Landau}: when the particle
is approaching the center of the potential, the kinetic energy $E_{k}=k^{2}/(2m_{\chi})$ with $k\sim r^{-1}$
increases as $1/r^{2}$ while $V$ decreases as $-1/r^{m}$.  Therefore,
the total energy would not be bounded from below and the particle
would keep falling to infinitely small $r$, corresponding to infinitely
high energy. Indeed, as mentioned above, for very small $r$ the
$1/r^{5}$ form will be no longer valid and the true form has lower
powers such as $1/r$ or $1/r^{2}$, depending on the UV behaviour
of the contact vertex. 

\begin{figure}
\centering

\includegraphics[width=0.5\textwidth]{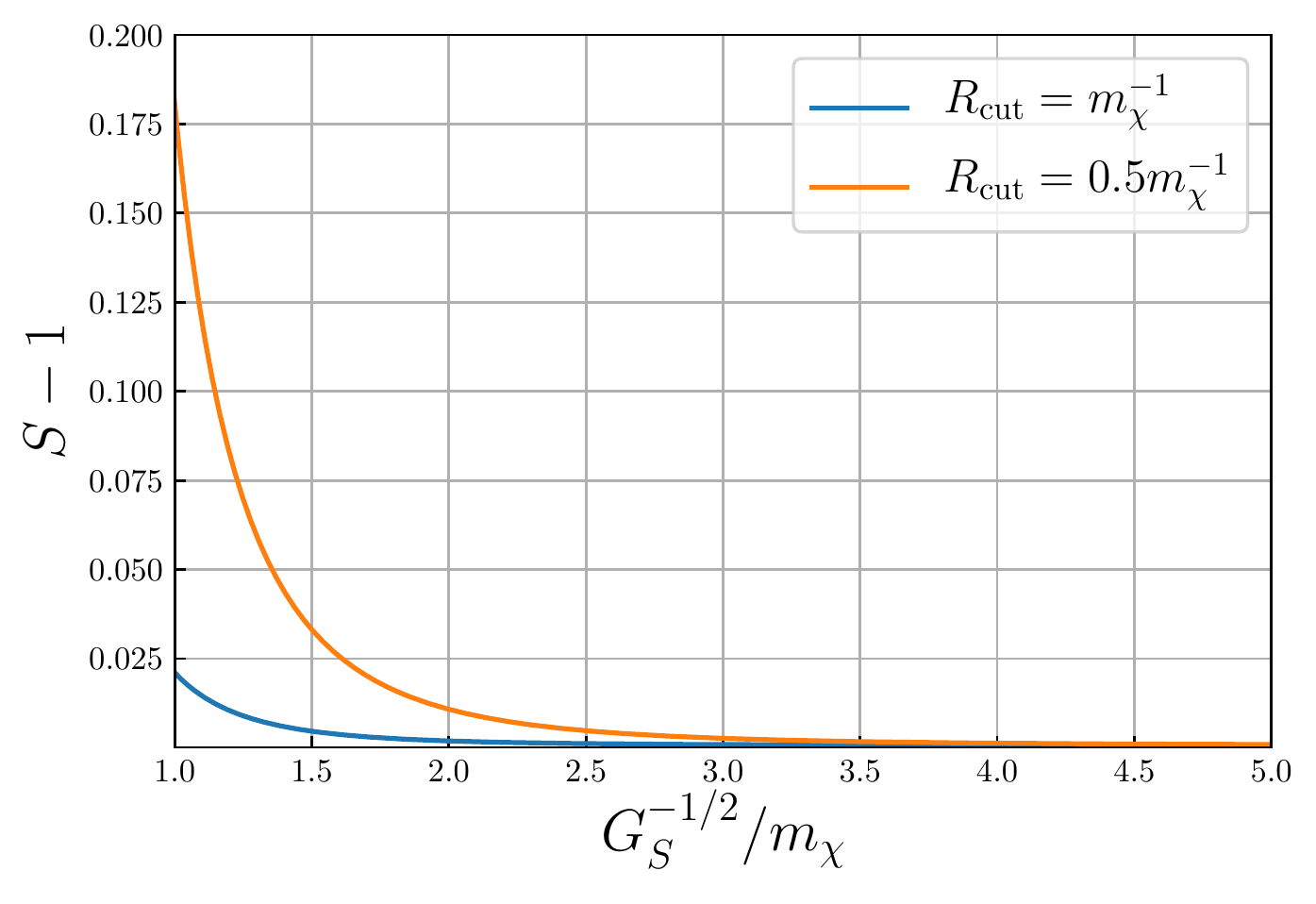}\includegraphics[width=0.49\textwidth]{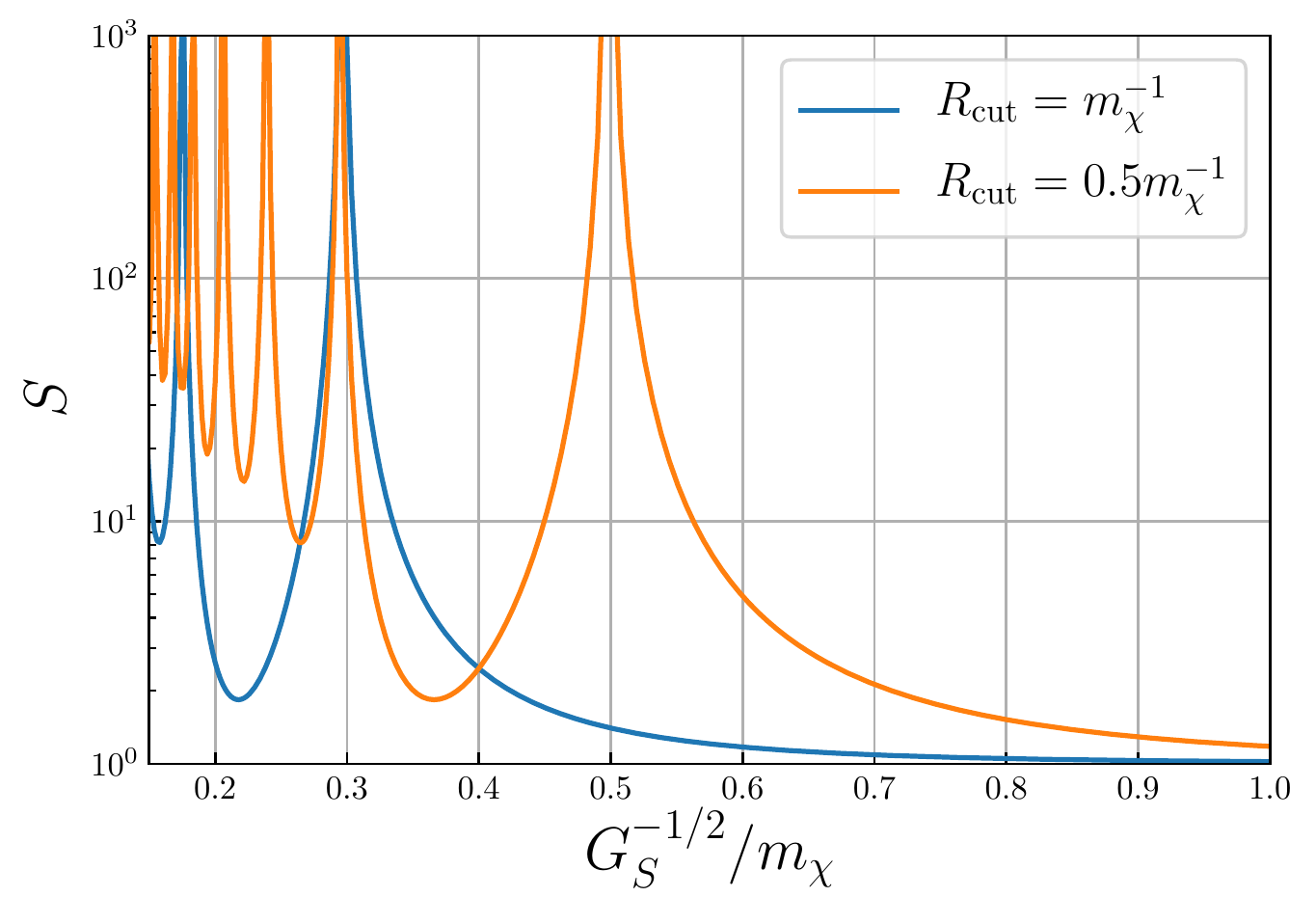}\caption{The Sommerfeld enhancement  due to  neutrinos forces derived from contact interactions. The left (right) panel shows a small (large) enhancement effect for $G_{S}^{-1/2}>m_\chi$ ($G_{S}^{-1/2}<m_\chi$). \label{fig:contact} 
}
\end{figure}

Despite the UV dependence, Eq.~\eqref{eq:V-con} is valid over a wide
range of large $r$  and we would like to ask whether in its valid
range it could cause significant Sommerfeld enhancement or not. To
this end, we introduce a  cut-off length scale $R_{{\rm cut}}$,
above which we employ Eq.~\eqref{eq:V-con} and below which we assume
the neutrino force vanishes (corresponding to a flat potential). 

By solving the Schr\"odinger equation, we obtain the Sommerfeld enhancement
factors in Fig.~\ref{fig:contact}. We set $R_{{\rm cut}}=0.5m_{\chi}^{-1}$
and $1.0m_{\chi}^{-1}$ since for $r\ll m_{\chi}^{-1}$ the non-relativistic
approximation would fail. In addition, the velocity is set by $k/m_{\chi}=10^{-3}$. 

Let us first focus on the left panel, where we show the values of
$S-1$ for $G_{S}^{-1/2}\geq1.0m_{\chi}$. In this regime, the $S$ factor
monotonically increases as the interaction strength $G_{S}$ increases.
The reason is obvious: a larger $G_{S}$ leads to a stronger neutrino
force and hence a stronger Sommerfeld enhancement effect. The result
depends on the cut-off scale, $R_{{\rm cut}}$. For $R_{{\rm cut}}=0.5m_{\chi}^{-1}$
($1.0m_{\chi}^{-1}$), $S-1$ cannot exceed $0.18$ ($0.021$) if $G_{S}^{-1/2}\geq1.0m_{\chi}$. 

One might ask what would happen if one further increases $G_{S}$. In the right
panel of Fig.~\ref{fig:contact}, we extend it to $G_{S}^{-1/2}\in[0.15,\ 1.0]m_{\chi}$.
As is shown in this plot, the $S$ factor can be drastically enhanced up to $S \gtrsim 10^3$
when $G_{S}^{-1/2}<1.0m_\chi$. In particular, the curves develop several
resonances, which is the typical behaviour of the Sommerfeld enhancement. 
Note, however, for $G_{S}^{-1/2}$ comparable to or less than $m_\chi$,
the $1/r^{5}$ form of the potential is likely to lose its validity
 before the non-relativistic approximation fails, because at $r\sim m_\chi^{-1}$,
the momentum transfer would be comparable to or greater than $G_{s}^{-1/2}$.
Therefore, the results obtained with $G_{S}^{-1/2}\lesssim1.0m_\chi$
are presented only for illustration: the correct answer depends the UV physics underlying the contact interaction.

\subsection{The $t$-channel case}
\label{subsec:t-channel}


 As discussed above, for potentials blowing up faster than $1/r^{2}$
(e.g.~$V\propto1/r^{m}$ with $m>2$), there is a theoretical inconsistency:
the particle would fall towards small $r$ with its kinetic
energy increasing infinitely. In Sec.~\ref{subsec:Contact}, we simply imposed a cut-off on the potential. In fact, sufficiently high energies would open the contact vertex, leading to an altered form of the potential. 

Let us now consider that the contact interaction is generated by a
$t$-channel mediator, with the Lagrangian given as follows:  
\begin{equation}
{\cal L}\supset - y_{\nu}\overline{\nu}\nu\phi - y_\chi\overline{\chi}\chi\phi-\frac{1}{2}m_{\phi}^{2}\phi^{2}\thinspace.\label{eq:L-t}
\end{equation}
At energies well below $m_{\phi}$, it generates the effective interaction
in Eq.~\eqref{eq:L-con} with 
\begin{equation}
G_{S}=\frac{y_{\nu}y_\chi}{m_{\phi}^{2}}\thinspace.\label{eq:-11}
\end{equation}
Note that at tree level, the mediator $\phi$ already induces an attractive
potential as formulated in Eq.~\eqref{eq:-12}. Hence the neutrino
force can alternatively be viewed as a special loop correction to
the tree-level potential. From Ref.~\cite{Xu:2021daf}, the effective
potential including  the neutrino force reads:
\begin{equation}
V_{t}\left(r\right)=\alpha_\chi\left[-\frac{e^{-m_{\phi}r}}{r}+\frac{\alpha_{\nu}}{4\pi}m_{\phi}\mathscr{V}\left(m_{\phi}r\right)\right],\label{eq:}
\end{equation}
where $\alpha_{\chi,\nu}\equiv y_{\chi,\nu}^{2}/(4\pi)$, 
\begin{equation}
\mathscr{V}\left(x\right)\equiv\frac{2+\left(2+x\right)e^{x}{\rm Ei}\left(-x\right)+\left(2-x\right)e^{-x}{\rm Ei}\left(x\right)}{x}\;,\label{eq:V-t}
\end{equation}
and ${\rm Ei}(x)\equiv-\int_{-x}^{\infty}z^{-1}e^{-z}dz$ is the exponential
integral function. As one can check, in the long-range limit ($r\gg m_{\phi}$),
the above potential returns to the $1/r^{5}$ form in Eq.~\eqref{eq:V-con},
whereas in the short-range limit ($r\ll m_{\phi}$) it behaves as
$1/r$, which implies that the Schr\"odinger equation can be solved
consistently without manually imposing any cut-off. 

\begin{figure}
\centering

\includegraphics[width=0.47\textwidth]{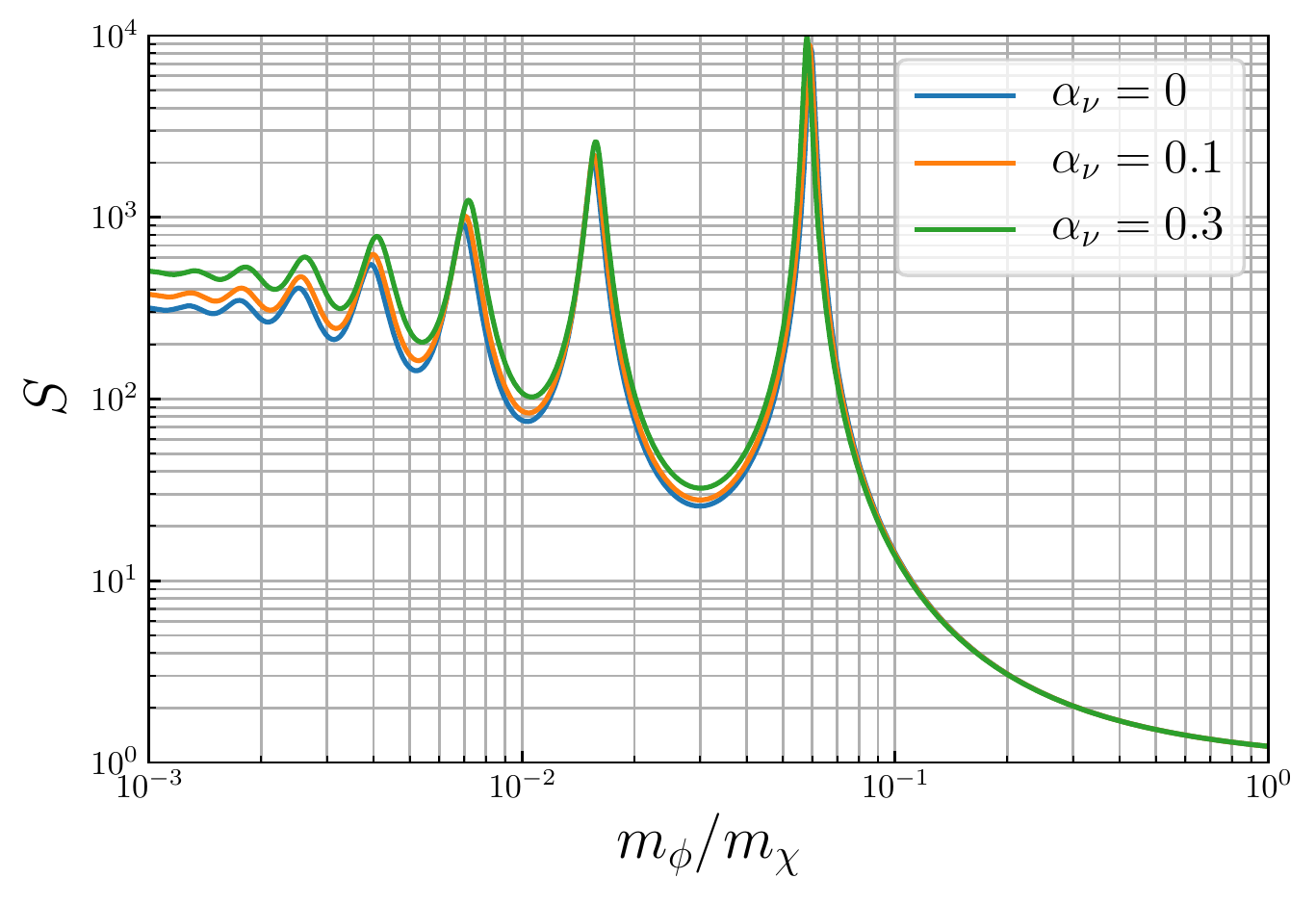}\includegraphics[width=0.49\textwidth]{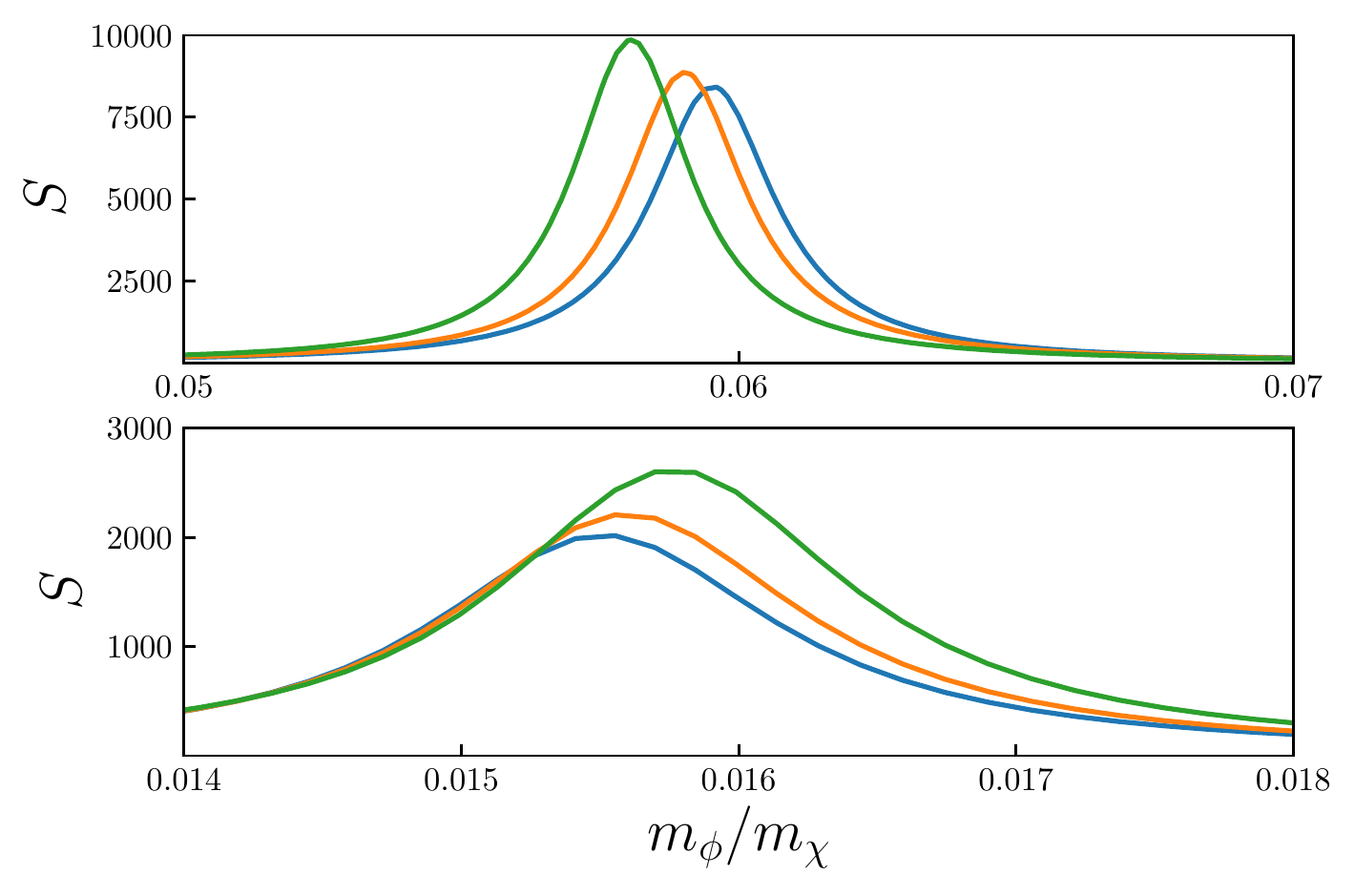}
\caption{The Sommerfeld enhancement factor $S$ as a function of the $t$-channel
mediator mass $m_{\phi}$. The  effective potential used to obtain
these curves is given in Eq.~\eqref{eq:V-t}, assuming $\alpha_{\nu}=\{0,0.1,0.3\}$
 and $\alpha_\chi=0.1$. The left panel shows the entire mass range
whereas the right panels focus on the two highest peaks. \label{fig:sommer-t} 
}
\end{figure}

Solving the Schr\"odinger equation with the potential in Eq.~\eqref{eq:V-t}
give the results displayed in Fig.~\ref{fig:sommer-t}.
Here we set $k/m_\chi=10^{-3}$, $\alpha_\chi=0.1$, and $\alpha_{\nu}=\{0,\ 0.1,\ 0.3\}$.
The left panel of Fig.~\ref{fig:sommer-t} shows how $S$ varies within $m_{\phi}\in[10^{-3},\ 1]m_\chi$, while the right panels focus
on two highest peaks, which are around $m_{\phi}\approx0.06m_\chi$ and $m_{\phi}\approx0.016m_\chi$.
The curves in Fig.~\ref{fig:sommer-t} can be viewed as the correct extrapolation of the curves in Fig.~\ref{fig:contact} for large $G_S$, assuming a $t$-channel UV completion of the contact interaction.

From the perspective of Feynman diagrams, the neutrino force in the
$t$-channel scenario is only a loop correction to the Yukawa force.
To inspect the role of neutrinos in the Sommerfeld enhancement, one
should compare the curves with $\alpha_{\nu}\neq0$ to the one with
$\alpha_{\nu}=0$. As shown in the left panel, the neutrino force
generally enhances the $S$ factor. For the parameters mentioned above,
we find that the peaks are enhanced by, from right to left (i.e. larger $m_\phi/m_\chi$ to smaller),
\begin{equation}
\frac{\Delta S}{S}=\begin{cases}
5.4\%,\ 9.5\%,\ 12\%,\ 14\%,\ \ldots,\ 19\% & {\rm for\ }\alpha_{\nu}=0.1\\
17\%,\ \ 29\%,\ \ 37\%,\ 43\%,\ \ldots,\ 60\% & {\rm for\ }\alpha_{\nu}=0.3
\end{cases}\thinspace,\label{eq:-17}
\end{equation}
where the last values are the small-$m_{\phi}$ limits of $\Delta S/S$.
This is much larger than usual one-loop corrections which are typically
of the order of $\alpha_{\nu}/(4\pi)=0.8\%$ or $2\%$ for $\alpha_{\nu}=0.1$
or 0.3 respectively. 
We have verified that varying $\alpha_\chi$ can increase or decrease the number and heights of the peaks but the relative ratios in Eq.~\eqref{eq:-17} remain rather stable with such changes. The results are also independent of $m_\chi$, as long as $m_\phi/m_\chi$ is fixed.

\subsection{The $s$-channel case}
\label{subsec:s-channel}
Another way to open the contact vertex is via an $s$-channel mediator.
To investigate how neutrino forces may affect the Sommerfeld enhancement
in this case, we consider the following Lagrangian: 
\begin{equation}
{\cal L}\supset -\left(y_{\nu}\overline{\chi}\nu\phi+{\rm h.c.}\right)-\frac{1}{2}m_{\phi}^{2}\phi^{2}\thinspace,\label{eq:L-s}
\end{equation}
 which  leads to the box diagram in Fig.~\ref{fig:feyn}. As has
been pointed out in Ref.~\cite{Xu:2021daf}, the effective potential
generated in this case contains spin-dependent and spin-independent pieces.
For simplicity, we focus on the latter which reads~\cite{Xu:2021daf}:
\begin{equation}
V_{s}\left(r\right)=-\frac{3\alpha_{\nu}^{2}}{8\pi r}\int_{0}^{\infty}\left[\frac{1}{2A}+\frac{B^{2}-tA}{4AB\sqrt{tA}}{\rm ln}\left(\frac{B-\sqrt{tA}}{B+\sqrt{tA}}\right)\right]e^{-\sqrt{t}r}dt\thinspace,\label{eq:V-s}
\end{equation}
where 
\begin{equation}
A\equiv t-4m_\chi^{2}\thinspace,\ B\equiv t+2m_{\Delta}^{2}\thinspace,\ m_{\Delta}^{2}\equiv m_{\phi}^{2}-m_\chi^{2}\thinspace.\label{eq:-16}
\end{equation}
In the long-range limit, Eq.~\eqref{eq:V-s} returns to the $1/r^{5}$
form whereas in the short-range limit, it becomes proportional to
$1/r^{2}$. More specifically, we have the following analytic approximations:
\begin{equation}
V_{s}\left(r\right)\approx-\frac{3\alpha_{\nu}^{2}}{8\pi}\times\begin{cases}
\frac{\pi}{4m_\chi r^{2}} & \ \ m_\chi^{-1}\ll r\ll m_{\Delta}^{-1}\\[2mm]
\frac{\pi}{4m_\chi m_{\Delta}^{2}r^{4}} & \ \ m_{\Delta}^{-1}\ll r\ll m_\chi m_{\Delta}^{-2}\\[2mm]
\frac{1}{m_{\Delta}^{4}r^{5}} & \ \ m_\chi m_{\Delta}^{-2}\ll r
\end{cases}\thinspace.\label{eq:-18}
\end{equation}

\begin{figure}
\centering

\includegraphics[width=0.5\textwidth]{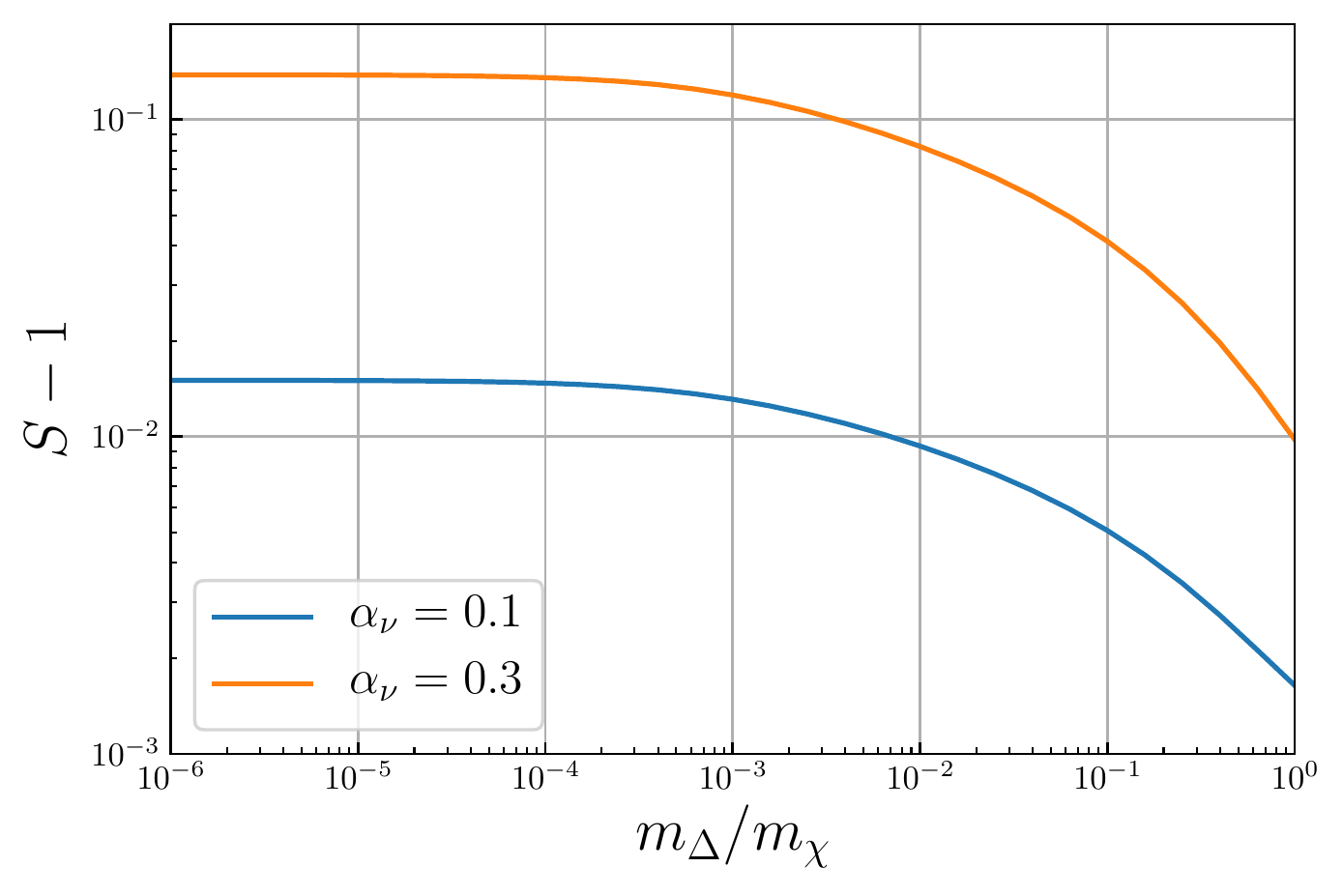}
\caption{The Sommerfeld enhancement factor $S$ in the $s$-channel case as
a function of $m_{\Delta}$ which is defined by  $m_{\Delta}^{2}\equiv m_{\phi}^{2}-m_\chi^{2}$.
\label{fig:sommer-s} 
}
\end{figure}

In fact, when $r$ is approaching $m_\chi^{-1}$, the potential behaves as $1/r^{\alpha}$ with $\alpha<2$, which justifies the use of quantum mechanics to compute the Sommerfeld enhancement. 
By directly substituting Eq.~\eqref{eq:V-s} into the Schr\"odinger
equation and solve it, we obtain the 
Sommerfeld enhancement factor presented in Fig.~\ref{fig:sommer-s}. 

According to Eq.~\eqref{eq:-18}, the most important mass scale in the $s$-channel case is the mass
squared difference $m_{\Delta}^{2}$ defined in Eq.~\eqref{eq:-16}.
A smaller $m_{\Delta}$ leads to a stronger neutrino force and hence a larger $S$, as is shown in Fig.~\ref{fig:sommer-s}. 
Overall, the Sommerfeld enhancement is not significant in the $s$-channel
case. For $\alpha_{\nu}=0.1\sim0.3$, the maximal enhancement is around
$1\%\sim10\%$. 

Similar to our previous discussion in the $t$-channel case, Fig.~\ref{fig:sommer-s}
can also be viewed as another large-$G_{S}$ extrapolation of the
curves in Fig.~\ref{fig:contact}. Unlike the $t$-channel case,
the $s$-channel extrapolation is flat without any resonances.


\section{Dark matter model-building and phenomenology}
\label{sec:DM pheno}
We have now seen that a neutrino force in the dark sector via a $t$-channel interaction from Eq.~\eqref{eq:L-t} can produce potentially sizeable modifications to the Sommerfeld enhancement.  
So far, we have been dealing with a toy model, since Eq.~\eqref{eq:L-t} is clearly not gauge-invariant and the nature of the dark matter particle $\chi$ is also unspecified. 
In this section, we will briefly discuss some of the model-building challenges and possible ways to avoid them. 
We will then consider the impact of this neutrino force on dark matter phenomenology.

First we note that the large $\alpha_\nu$ and $\alpha_\chi$ required to modify the Sommerfeld enhancement implies that both the $\phi$ and $\chi$ will equilibrate with the SM. 
Consequently, $m_\chi, m_\phi \gtrsim$ MeV is required to avoid affecting BBN too greatly, given $\Delta N_{\rm eff} \lesssim 0.3$ \cite{Fields:2019pfx}.

The Yukawa interaction $\overline{\nu} \nu \phi$ could involve a) only left-handed neutrinos, in which case we have $\overline{\nu_L^c} \nu_L \phi + {\rm h.c.}$, b) a left-handed and a right-handed neutrino, $\overline{\nu_L} \nu_R \phi + {\rm h.c.}$, or c) only right-handed neutrinos, $\overline{\nu_R^c} \nu_R \phi + {\rm h.c.}$. 
Then, the requirement of gauge invariance implies that either $\alpha_\nu$ or $\alpha_\chi$ should inevitably receive some type of suppression.

If the scalar couples to at least one left-handed neutrino, then it must have a non-trivial representation under $SU(2)_L \times U(1)_Y$. 
The only gauge-invariant renormalisable term which induces a $\overline{\nu_L^c} \nu_L \phi$ interaction is $\overline{l_L^c} i\sigma_2 \sigma^a \Delta^a l_L + {\rm h.c.}$, with $\Delta^a \sim \mathbf{3}_1$, where the first number is the representation under $SU(2)_L$ and the second is the hypercharge. 
Similarly, the $\overline{\nu_L} \nu_R \phi$ interaction is generated at the renormalisable level by a term of the form $\overline{l_L} \tilde{\Phi} \nu_R + {\rm h.c.}$, where $\Phi \sim \mathbf{2}_{1/2}$. 
The first consequence of a scalar charged under the EW symmetry is that 
its mass has to be high to avoid collider bounds\footnote{For the triplet $\Delta$ which contains a doubly charged Higgs, the latest bound lies between 500 to 800 GeV, depending on the lepton flavour~\cite{CMS-PAS-HIG-16-036}. For the doublet $\Phi$ which is mainly constrained by its singly charged Higgs, the latest bound is around $80$ GeV~\cite{Abbiendi:2013hk}.  }.
The second is that a $\overline{\chi} \chi \phi$ coupling implies that $\chi$ should be the neutral component of a low-dimensional $SU(2)_L$ multiplet. 
However, direct detection rules out such DM candidates, see e.g.~\cite{Cirelli:2005uq}. 
The DM can be made sterile with respect to the SM via various mechanisms, for instance if the $\phi$ mixes with some neutral scalar that couples to $\chi$, or if the $\overline{\chi} \chi \phi$ interaction is generated from some higher-dimensional operator. 
These mechanisms generically face a suppression factor. 
The mixing angle, for instance, scales as $\theta \sim \langle \phi \rangle/m_\phi \lesssim \mathcal{O}(10^{-2})$, since the VEV of an EW-charged scalar is constrained to be $\langle \phi \rangle \lesssim $ GeV by EW precision data~\cite{ParticleDataGroup:2020ssz}.

The scalar may be 
a singlet without SM charges  
if it couples only to right-handed neutrinos, which is reminiscent of the Majoron model~\cite{Chikashige:1980ui}. Motivated by the Majoron model, it could be a pseudoscalar which however would lead to suppressed neutrino forces, as previously mentioned. 
If we consider a scalar instead of a pseudoscalar, then the $\overline{\nu_R^c} \nu_R \phi$ term can have an $\mathcal{O}(1)$ coupling\footnote{
	The experimental constraints on such a scenario are very weak since $\phi$ does not directly couple to SM particles and its loop-induced couplings to SM particles is highly suppressed by neutrino masses~\cite{Xu:2020qek}.}
and hence significant Sommerfeld enhancements, provided that  $\nu_R$ is much lighter than the DM.
However, in this case the DM phenomenology is most likely limited to the dark sector because the active-sterile neutrino mixing must be less than $\mathcal{O}(10^{-2})$~\cite{Fernandez-Martinez:2016lgt}. 

\subsection{Beyond scalar-mediated neutrino force}

\begin{figure}
\centering

\includegraphics[width=0.8\textwidth]{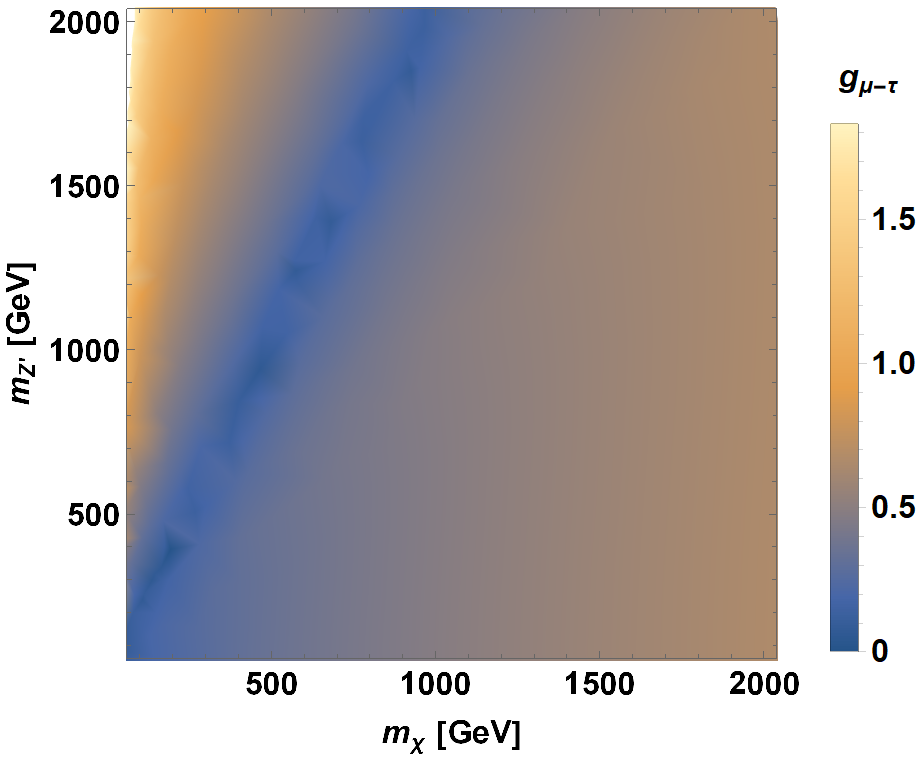}
\caption{Masses and gauge coupling required for correct DM abundance in the $L_{\mu} - L_{\tau}$ model. \label{fig:Zp}}
\end{figure}

The requirement of gauge invariance is one of the main challenges for model-building with a scalar mediator. 
The situation is much simpler, however, when considering a $Z'$ model. 
The $Z'$ of a gauged $L_{\mu} - L_{\tau}$ is permitted by experiments to have order-one couplings as long as its mass 
is $m_{Z'} \gtrsim 200$ GeV, around which it is mainly constrained by the muon $g-2$ and CCFR---see e.g.~\cite{Coy:2021wfs}.
One can then introduce a vector-like fermion (thereby avoiding gauge anomalies), $\chi$, which is charged under the $L_{\mu} - L_{\tau}$, as the DM candidate. 
The resulting interactions,
\begin{equation}
    \mathcal{L} \supset g_{\mu - \tau} Z'_\alpha (\overline{\nu_\mu} \gamma^\alpha P_L \nu_\mu - \overline{\nu_\tau} \gamma^\alpha P_L \nu_\tau + \overline{\chi} \gamma^\alpha \chi) + \ldots \, , \label{eq:L-tVector}
\end{equation}
form a vectorial version of Eq.~\eqref{eq:L-t}. 
Above, the ellipsis is for the $Z'$ coupling to the mu and tau. 
The DM freezes out via $s$-channel annihilations into charged leptons and neutrinos, and $t$- and $u$-channel annihilations into $Z'$ pairs. 
As shown in Fig.~\ref{fig:Zp}, this scenario gives the correct DM abundance for $\mathcal{O}(1)$ gauge coupling, except in the band $m_{Z'} \simeq 2m_\chi$, where there is an $s$-channel resonance. 
The neutrino forces arising in the $t$-channel mediated by the vector boson can be calculated in the same manner as 
the scalar case in \cite{Xu:2021daf}. The resulting potentials have a similar $r$-dependence to the scalar case, though the spin dependence is different and more complicated.
We leave a dedicated study of the vector case to future work.

Beyond this, we point out that any fermion sufficiently light compared to both the DM and mediator can replicate the effect of the neutrino force on the Sommerfeld enhancement. 
One may therefore equally have an electron (or muon or tau) force when both DM and the mediator are much heavier than its mass scale. 
Not only does this open up further model-building possibilities, it could also perhaps have greater phenomenological consequences, since generally DM-electron interactions are better constrained than DM-neutrino ones.

\subsection{Dark matter phenomenology}
The Sommerfeld enhancement may be important for a number of aspects of DM phenomenology, as outlined in the introduction. 
Here we briefly survey the implications of dark matter whose Sommerfeld enhancement is modified due to the presence of a neutrino force.

Firstly, a DM-neutrino interaction implies the possibility of indirect detection of DM via $\chi \overline{\chi} \to \nu \overline{\nu}$ annihilations. 
The experimental bounds on this process remain much weaker than on DM annihilations into photons or charged particles. 
Indeed, an analysis of constraints for dark matter masses from MeV to ZeV scales found that except for a narrow band around $m_\chi \sim 10-30$ MeV, the thermally-averaged cross section is allowed to be several orders of magnitude larger than the one required for the correct relic abundance in the canonical freeze-out scenario, $\langle \sigma v \rangle \simeq 3 \times 10^{-26} \text{cm}^3/\text{s}$ \cite{Arguelles:2019ouk}. 
However, this channel may be important if neutrinos are the only SM particles with which DM interacts; alternatively it may provide a signal which, in conjunction with others, would allow us to determine more precisely the nature of DM. 
Given the weakness of the bounds, a Sommerfeld enhancement factor of up to $10^4$ (cf. Fig.~\ref{fig:sommer-t}) would be a neat explanation of how DM could produce an observable neutrino signature yet still be consistent with the correct relic abundance. 
A neutrino force which increases this Sommerfeld factor by up to $\sim 60\%$ further opens up this possibility. 
Note also that, as mentioned above, an equivalent `electron force' can be realised if both the DM and mediator masses are much heavier than the electron. 
In this case, such a force would further enhance the rate of $\chi \overline{\chi} \to e^+ e^-$ compared to the typical expectation, which could potentially leave a far more striking signature. 

Secondly, it has been remarked that the Sommerfeld enhancement affects the standard freeze-out calculation and can modify the relic abundance by an order-one factor \cite{Hisano:2006nn}. 
This would be additionally affected in the presence of a neutrino force, and therefore we remark that a precise freeze-out calculation for DM which interacts with neutrinos requires careful consideration of the neutrino force. 

DM self-interactions are also known as a possible resolution of tensions between cold DM predictions and observations in small-scale structures \cite{Tulin:2017ara}. 
In particular, self-interactions which have a Sommerfeld enhancement at low velocities are particularly attractive as they can facilitate $\sigma/m \sim 1\text{cm}^2/\text{g}$ required to ameliorate the tensions while avoiding constraints from observations at higher DM velocities, where the enhancement is less effective. 
If DM experiences a substantial neutrino force, not only will its annihilations into neutrinos be further enhanced, but also its own scatterings.

\section{Conclusion }
\label{sec:summary}
The Sommerfeld enhancement is by now an established part in dark matter phenomenology. 
In this paper we went beyond the canonical light-mediator scenario and investigated the role of a neutrino force on this effect.

DM-neutrino interactions should be considered at two scales, depicted in Fig.~\ref{fig:feyn}. After outlining the formalism in Sec.~\ref{sec:SE}, we began Sec.~\ref{sec:NuForce} by considering the long-range force generated by the exchange of a pair of neutrinos, which produces a potential of the form $V \propto 1/r^5$. We showed that within the region of validity of this potential, it induces only a very modest Sommerfeld enhancement, as displayed in the left panel of Fig.~\ref{fig:contact}. 
At shorter scales, the UV completion must be considered. 
We computed the Sommerfeld enhancement for two potentials valid at short distances recently derived in \cite{Xu:2021daf}, and saw that the results varied significantly between the different cases. 
While the enhancement factor is negligible in the $s$-channel case, the $t$-channel one can induce an enhancement as large as $\mathcal{O}(10^4)$. 
Importantly, although this effect is generated by the DM interaction with the scalar mediator alone, the presence of the neutrino interaction can modify it by as much as $\mathcal{O}(60\%)$. 
This is much greater than what would be expected of a loop correction, and is one of the main results of this paper.

Our results are rather general and can be applied to any instance in which the DM interacts with a light fermion in a form given by Eqs.~\eqref{eq:L-con}, \eqref{eq:L-t} or \eqref{eq:L-s}. 
The critical point of distinction between the two UV completions is that at short distances the $t$-channel potential scales as $V \propto 1/r$ while for the $s$-channel it behaves as $V \propto 1/r^2$. 
The behaviour of the DM wave function for these and more general potentials was studied in Appendix~\ref{sec:Small-r}.

Finally, we commented on model-building and dark matter phenomenology in Sec.~\ref{sec:DM pheno}. 
Although it is challenging to construct a satisfactory $t$-channel model with a scalar mediator, we demonstrated that the basic idea fits into a greater class of models. The mediator may be a vector, as illustrated in the example of a gauged $L_\mu - L_\tau$. Alternatively, for sufficiently heavy dark matter, the effect of the neutrino force may be replicated by an electron (or muon or tau) force instead. 
There are a variety of implications, including for indirect detection experiments and freeze-out calculations, which warrant further investigation.

\section*{Acknowledgments}
This project has received support from the IISN convention 4.4503.15 and from 
the National Natural Science Foundation of China under grant No. 12141501 and No. 11835013.

\appendix

\section{Small-$r$ behaviour of radial wave functions in inverse-power potentials\label{sec:Small-r}}




In this appendix, we perform a systematic study of the short-range behaviour of the radial wave function in general inverse-power potentials. 
The radial Schr\"odinger equation reads
\begin{eqnarray}
u''(r)+\left[2\mu\left(E-V\right)-\frac{l\left(l+1\right)}{r^2}\right]u(r)=0\;,
\end{eqnarray}
where $E=\mu v^2/2>0$ is the energy of relative motion, with $v$ the relative velocity between two particles. Defining the dimensionless quantity $\rho\equiv \sqrt{2\mu E}\,r$, one obtains
\begin{eqnarray}
u''(\rho)+\left[1-\frac{V(\rho)}{E}-\frac{l\left(l+1\right)}{\rho^2}\right]u(\rho)=0\;.
\end{eqnarray}
Suppose we have an inverse-power potential like
\begin{eqnarray}
V(\rho)=-\frac{m_0}{\rho^d}\;,
\end{eqnarray}
where $m_0^{}$ is a parameter with the dimension of mass and $d$ is an arbitrary positive number. Then the radial equation is
\begin{eqnarray}
	\label{eq:general radial equation}
u''(\rho)+\left[1+\frac{b}{\rho^d}-\frac{l\left(l+1\right)}{\rho^2}\right]u(\rho)=0\;,
\end{eqnarray}
with $b\equiv m_0^{}/E$. It is obvious that $b>0$ ($b<0$) corresponds to an attractive (repulsive) potential. The physical initial condition of the radial wave function requires $\lim_{\rho\to0}u(\rho)=0$. However, it can be shown that for $d>2$ and $b>0$, the solution of Eq.~(\ref{eq:general radial equation}) will not tend to zero near the origin~\cite{Frank:1971xx}. More precisely, a potential $V(\rho)$ that satisfies 
\begin{eqnarray}
\lim_{\rho\to0}\rho^2V(\rho)=0\;\;\;({\rm or} \pm \infty)
\end{eqnarray}
is called \emph{regular} (or \emph{singular}) at $\rho=0$, while a potential satisfying 
\begin{eqnarray}
\lim_{\rho\to0}\frac{\log V(\rho)}{\log \rho}=-2
\end{eqnarray}  
is said to be a \emph{transitional} potential. For regular or transitional potentials, the solution of Eq.~(\ref{eq:general radial equation}) can be expanded as a series of $\rho$ near $\rho=0$ (using the Frobenius method discussed below), while for a singular potential there does not exist a series solution near the origin. 
In fact, the solutions of Eq.~(\ref{eq:general radial equation}) with singular potential have the following asymptotic behaviour as $\rho\to0$~\cite{Frank:1971xx}:
\begin{eqnarray}
	\label{eq:asymptotic behaviour}
u(\rho)\sim \left|V(\rho)\right|_{}^{-1/4}{\rm exp}\left(\pm i \int^{\rho} \left[-V(\rho^\prime)\right]^{1/2}d\rho^\prime\right)\;.
\end{eqnarray}
In particular, for an attractive singular potential ($b>0$, $d>2$), Eq.~(\ref{eq:asymptotic behaviour}) shows that the solution will oscillate infinitely rapidly as it tends to the origin. This corresponds to an unacceptable solution 
since it does not have a unique bound-state spectrum, has an infinite number of bound states and no lower bound on the energy~\cite{Frank:1971xx}. On the contrary, a repulsive singular potential ($b<0$, 
$d>2$) can lead to a solution which decreases to zero exponentially near the origin~\cite{Case:1950an,giudice1965singular,giudice1965singular2}
\begin{eqnarray}
u(\rho)\sim \rho^{d/4}{\rm exp}\left(-\frac{2}{d-2}\, \rho^{-\frac{d-2}{2}}\right)\;,
\end{eqnarray}
which is well-defined for physical problems. 
Throughout this paper, we will only focus on the short-range behaviour of the solutions for regular and transitional potentials. This is because the $t$- and $s$-channel case of neutrino forces behave as $1/r$ and $1/r^2$ respectively at small distance~\cite{Xu:2021daf}, both of which are well-behaved (i.e.~not singular) potentials. In the next part of this appendix, we will introduce the general method to obtain the series solutions near the origin for regular and transitional potentials.

\subsection{Frobenius method}
Consider the following second-order linear differential equation for $u(\rho)$:
\begin{eqnarray}
	\label{eq:general 2nd diff}
u''(\rho)+P(\rho)u'(\rho)+Q(\rho)u(\rho)=0\;,
\end{eqnarray}
where $P(\rho)$ and $Q(\rho)$ are some general functions. 
If $P(\rho)$ or $Q(\rho)$ has a pole at $\rho=0$, then one cannot simply expand the solution as a series at $\rho=0$ because it is the singular point of this equation. However, the \emph{theorem of Frobenius} shows that if both $\tilde{P}(\rho)\equiv\rho\,P(\rho)$ and $\tilde{Q}(\rho)\equiv\rho^2Q(\rho)$ are finite as $\rho$ approaches zero\footnote{In this case, $\rho=0$ is called a \emph{regular} singular point of the equation, otherwise it is an \emph{irregular} singular point. It is obvious that for regular and transitional potentials with $d\leqslant 2$ in Eq.~(\ref{eq:general radial equation}), $\rho=0$ is a regular singular point, while for singular potential with $d>2$, $\rho=0$ is an irregular singular point.}, then one can still obtain the general series solution~\cite{Teschlordinary2012}
\begin{eqnarray}
	\label{eq:general series solution}
u(\rho)=\rho^\nu\sum_{n=0}^{\infty}a_n\rho^n\;,
\end{eqnarray}
where $\nu$ is some number to be determined. Note that one can always adjust the value of $\nu$ to make $a_0\neq0$. In this case, $\nu$ is determined by the \emph{indicial equation},
\begin{eqnarray}
\nu\left(\nu-1\right)+\tilde{P}(0)\nu+\tilde{Q}(0)=0\;,
\end{eqnarray}
where $\tilde{P}(0)$ and $\tilde{Q}(0)$ are the limits of $\tilde{P}(\rho)$
and $\tilde{Q}(\rho)$ as $\rho\to 0$. Below we will illustrate how to apply the Frobenius method through some concrete examples.

\subsection{Coulomb potential}
To start, we consider the Coulomb potential, which corresponds to $d=1$ in Eq.~(\ref{eq:general radial equation}). Substituting Eq.~(\ref{eq:general series solution}) into Eq.~(\ref{eq:general radial equation}), one obtains
\begin{eqnarray}
&&a_0\rho^\nu\left[\nu\left(\nu-1\right)-l\left(l+1\right)\right]+\rho^{\nu+1}\left\{a_1\left[\nu\left(\nu+1\right)-l\left(l+1\right)\right]+b\,a_0\right\}\nonumber\\
&+&\sum_{n=2}^{\infty}\rho^{\nu+n}\left\{a_n\left[\left(n+\nu\right)\left(n+\nu-1\right)-l\left(l+1\right)\right]+a_{n-2}+b\,a_{n-1}\right\}=0\;.
\end{eqnarray}
Considering the coefficient of $a_0$, with $a_0 \neq 0$ by convention, we have the indicial equation
\begin{eqnarray}
\nu\left(\nu-1\right)-l\left(l+1\right)=0\;,
\end{eqnarray}
which gives $\nu_1=l+1$ or $\nu_2=-l$.
Then the terms proportional to $\rho^{\nu+1}$ and $\rho^{\nu+n}$ give the recursive relation
\begin{eqnarray}
	\label{eq:recursive relation}
a_1=\frac{b\,a_0}{l\left(l+1\right)-\nu\left(\nu+1\right)}\;,\quad
a_n=\frac{-a_{n-2}-b\,a_{n-1}}{\left(n+\nu-1\right)\left(n+\nu\right)-l\left(l+1\right)}\;\; ({\rm for}\;n\geqslant2)\;.
\end{eqnarray}
For $l\geqslant 0$, the two roots of $\nu$ along with the recursive relation give two independent solutions, with the general solution being a linear combination of these,
\begin{eqnarray}
u_l(\rho)&=&c_1\rho^{l+1}\left[1-\frac{b}{2\left(l+1\right)}\rho+\frac{b^2-2\left(l+1\right)}{4\left(l+1\right)\left(2l+3\right)}\rho^2+\ldots\right]\nonumber\\
&+&c_2\rho^{-l}\left[1+\frac{b}{2l}\rho+\frac{b^2+2l}{4l\left(2l-1\right)}\rho^2+\ldots\right]\;,\quad l\geqslant 1\;,
\label{eq:chilsoln}
\end{eqnarray}
where $c_1$ and $c_2$ are two arbitrary numbers to be determined by initial conditions, and the ellipses denote terms with higher powers of $\rho$. It is obvious that terms proportional to $\rho^{-l}$ blow up when $\rho \to 0$, so we must set $c_2=0$. Thus, the asymptotic behaviour of $u_l(\rho)$ as $\rho$ approaches zero is
\begin{eqnarray}
u_l(\rho)\sim \rho^{l+1},\quad
l\geqslant1\;.
\end{eqnarray}
For the case of $l=0$, the second line of Eq.~\eqref{eq:chilsoln} blows up. Therefore, one should firstly take the solution taking with only the $\nu_1 = l+1$ root of the indicial equation, 
\begin{eqnarray}
u_0^{(1)}(\rho)=\rho\left(1-\frac{b}{2}\rho+\frac{b^2-2}{12}\rho^2+\ldots\right)\;.
\end{eqnarray}
Then the second particular solution cannot be deduced from the recursion relation in Eq.~(\ref{eq:recursive relation}). Rather, it can be calculated from the first particular solution by
\begin{eqnarray}
u_0^{(2)}(\rho)=u_0^{(1)}(\rho)\int^{\rho} \frac{{\rm exp}\left[-\int^{\rho'} P(\rho'')d\rho''\right]}{\left[u_0^{(1)}(\rho')\right]^2}d\rho'\;,
\end{eqnarray}
where $P(\rho)$ is the function 
defined in Eq.~(\ref{eq:general 2nd diff}) and is zero in this case. Thus, it is straightforward to calculate the second particular solution,
\begin{eqnarray}
u_{0}^{(2)}=b\,u_0^{(1)}(\rho)\log\rho-\left[1-\frac{b}{2}\rho-\frac{b^2+1}{2}\rho^2+\frac{b\left(11b^2+2\right)}{72}\rho^3+\ldots\right]\;,
\end{eqnarray}
and the general solution for $l=0$ is given by a linear combination of $u_0^{(1)}(\rho)$ and $u_0^{(2)}(\rho)$. However, in order to guarantee $\lim_{\rho\to0}u_0(\rho)=0$, the term proportional to $u_0^{(2)}(\rho)$ must vanish. Hence, the asymptotic behaviour of $u_0^{}(\rho)$ as $\rho\to0$ is
\begin{eqnarray}
u_0(\rho)\sim \rho\;.
\end{eqnarray}
The wave function is therefore dominated by the $l=0$ mode as $\rho\to0$,
\begin{eqnarray}
\psi(\rho)\sim \sum_{l=0}^{\infty}P_l\left(\cos\theta\right)\frac{u_l(\rho)}{\rho}\sim \frac{u_0(\rho)}{\rho}\sim \text{finite constant}\;.
\end{eqnarray}

\subsection{Regular potential}
Next we consider the general regular potential, where $0<d<2$ in Eq.~(\ref{eq:general radial equation}). For simplicity, we assume that $d$ is a rational number, so that it can be written as the ratio of two positive integers $d=s/t$ with $s<2t$. The radial equation reads
\begin{eqnarray}
u''(\rho)+\left[1+\frac{b}{\rho^{s/t}}-\frac{l\left(l+1\right)}{\rho^2}\right]u(\rho)=0\;.
\end{eqnarray}
The key observation is that one can change the variable from $\rho$ to $\tilde{\rho}=\rho^{1/t}$, then the radial equation turns out to be
\begin{eqnarray}
	\label{eq:regular}
\tilde{\rho}^2u''(\tilde{\rho})+\left(1-t\right)\tilde{\rho}u'(\tilde{\rho})+t^2\left[\tilde{\rho}^{2t}+b\tilde{\rho}^{2t-s}-l\left(l+1\right)\right]u(\tilde{\rho})=0\;.
\end{eqnarray}
Note that since $s<2t$, the term proportional to $u(\tilde{\rho})$ in Eq.~(\ref{eq:regular}) is finite and independent of $\tilde{\rho}$ as $\tilde{\rho}\to 0$. Thus the indicial equation is given by
\begin{eqnarray}
\nu\left(\nu-1\right)+\left(1-t\right)\nu-t^2l\left(l+1\right)=0\;,
\end{eqnarray}
with the solutions $\nu_1=t\left(l+1\right)$ and $\nu_2=-tl$. For $l\geqslant1$, the general solution can be written as 
\begin{eqnarray}
u_l({\rho})&=&c_1\tilde{\rho}^{t(l+1)}(1+a_1\tilde{\rho}+a_2\tilde{\rho}^2+\ldots)+c_2\tilde{\rho}^{-tl}(1+a_1^\prime\tilde{\rho}+a_2^\prime\tilde{\rho}^2+\ldots)\nonumber\\
&=&c_1\rho^{l+1}(1+a_1\rho^{1/t}+a_2\rho^{2/t}+\ldots)+c_2\rho^{-l}(1+a_1^\prime \rho^{1/t}+a_2^\prime\rho^{2/t}+\ldots)\,,
\end{eqnarray}
with $c_1$ and $c_2$ arbitrary numbers. The initial condition requires $c_2=0$, thus the asymptotic behaviour of $u_l(\rho)$ as $\rho\to0$ is
\begin{eqnarray}
u_l(\rho)\sim \rho^{l+1}\;,\quad
l\geqslant1\;.
\end{eqnarray}
As for $l=0$, again one should only use the solution from the first root of the indicial equation, $\nu_1=t\left(l+1\right)$, to obtain
\begin{eqnarray}
u_0^{(1)}\left(\tilde{\rho}\right)=\tilde{\rho}^{t}\left(1+a_1\tilde{\rho}+\ldots\right)\;,
\end{eqnarray} 
while the second particular solution is given by
\begin{eqnarray}
	u_0^{(2)}(\tilde{\rho})&=&u_0^{(1)}(\tilde{\rho})\int^{\tilde{\rho}} \frac{{\rm exp}\left[-\int^{\tilde{\rho}'} P(\tilde{\rho}'')d\tilde{\rho}''\right]}{\left[u_0^{(1)}(\tilde{\rho}')\right]^2}d\tilde{\rho}' \nonumber\\
	&=&u_0^{(1)}(\tilde{\rho})\log\tilde{\rho}+C\left(1+a_1^\prime \tilde{\rho}+\ldots\right)\;,
\end{eqnarray}
where $C$ and $a_1^\prime$ are some irrelevant numbers that do not influence the short-distance asymptotic behaviour of the wave function. Note that we have used $P\left(\tilde{\rho}\right)=\left(1-t\right)/\tilde{\rho}$ in this case. Then the general solution can be written as 
\begin{eqnarray}
u_0(\rho)=c_1u_0^{(1)}\left(\tilde{\rho}\right)+c_2u_0^{(2)}\left(\tilde{\rho}\right)=c_1u_0^{(1)}\left(\rho^{1/t}\right)+c_2u_0^{(2)}\left(\rho^{1/t}\right)\;.
\end{eqnarray}
The initial condition $\lim_{\rho\to0}u_0(\rho)=0$ enforces $c_2=0$, thus the asymptotic behaviour of $u_0(\rho)$ as $\rho\to 0$ reads
\begin{eqnarray}
u_0(\rho)\sim \tilde{\rho}^{t}\sim\rho\;,
\end{eqnarray}
while that of the wave function is given by
\begin{eqnarray}
\psi(\rho)\sim \sum_{l=0}^{\infty}P_l\left(\cos\theta\right)\frac{u_l(\rho)}{\rho}\sim \frac{u_0(\rho)}{\rho}\sim \text{finite constant}\;.
\end{eqnarray}
This completes the proof that for regular potentials, the asymptotic behaviour of the wave function at short distance is always dominated by the $l=0$ mode.

\subsection{Transitional potential}
\renewcommand\arraystretch{1.5}
\begin{table}[t!]
	\centering
	\begin{tabular}{c|c|c}
		\hline \hline
		values of $b$ &  dominant modes & asymptotic behaviour of radial wave function \\
		\hline \hline
		$b\leqslant0$ & $l=0$ & $\rho^{\frac{1}{2}\left(1+\sqrt{1-4b}\right)}$\\
		\hline
		$0<b\leqslant\frac{1}{4}$ & $l=0$ & $\rho^{\frac{1}{2}\left(1-\sqrt{1-4b}\right)}$\\
		\hline
		$\frac{1}{4}<b\leqslant2$ & $l=0$ & $\rho^{\frac{1}{2}\left(1\pm i\sqrt{4b-1}\right)}$\\
		\hline
		$2<b\leqslant\frac{9}{4}$ & $l=1$ & $\rho^{\frac{1}{2}\left(1-\sqrt{9-4b}\right)}$\\
		\hline
		$\frac{9}{4}<b\leqslant6$ & $l=0,1$ & $\rho^{\frac{1}{2}\left(1\pm i\sqrt{4b-1}\right)}$, $\rho^{\frac{1}{2}\left(1\pm i\sqrt{4b-9}\right)}$\\
		\hline
		$6<b\leqslant\frac{25}{4}$ & $l=2$ & $\rho^{\frac{1}{2}\left(1-\sqrt{25-4b}\right)}$\\
		\hline
		$\frac{25}{4}<b\leqslant12$ & $l=0,1,2$ & $\rho^{\frac{1}{2}\left(1\pm i\sqrt{4b-1}\right)}$, $\rho^{\frac{1}{2}\left(1\pm i\sqrt{4b-9}\right)}$,
		$\rho^{\frac{1}{2}\left(1\pm i\sqrt{4b-25}\right)}$\\
		\hline
		\hline
	\end{tabular}
	\vspace{0.5cm}
	\caption{Some values of the coupling constant $b$ for the transitional potential, along with the dominant modes at short ranges and the asymptotic behaviour of the radial wave function as $\rho\to0$. For the purpose of illustration, we only list values of $b$ no larger than 12.}
	\label{table}
	\renewcommand\arraystretch{1}
\end{table}

Finally, let us consider the case of the transitional potential, namely $d=2$ in Eq.~(\ref{eq:general radial equation}). The indicial equation reads
\begin{eqnarray}
	\label{eq:nu1nu2}
\nu\left(\nu-1\right)+b-l\left(l+1\right)=0\quad\Rightarrow\quad
\nu_{1,2}=\frac{1\pm\sqrt{\left(2l+1\right)^2-4b}}{2}\;.
\end{eqnarray}
Note that the indicial equation in the scenario of transitional potential depends on the coupling constant of the potential, $b$. This is the main difference compared to regular potentials. Using the Frobenius method, it is straightforward to obtain
\begin{eqnarray}
	\label{eq:general transitional function}
u_l(\rho)&=&c_1\rho^{\nu_1}\left[1-\frac{1}{2\left(1+2\nu_1\right)}\rho^2+\frac{1}{8\left(1+2\nu_1\right)\left(3+2\nu_1\right)}\rho^4+\ldots\right]\nonumber\\
&+&c_2\rho^{\nu_2}\left[1-\frac{1}{2\left(1+2\nu_2\right)}\rho^2+\frac{1}{8\left(1+2\nu_2\right)\left(3+2\nu_2\right)}\rho^4+\ldots\right]\;,
\end{eqnarray}
where $\nu_1$ and $\nu_2$ are given by Eq.~(\ref{eq:nu1nu2}), while $c_1$ and $c_2$ are two arbitrary numbers. In contrast to the regular potential, the dominant mode at short distance in Eq.~(\ref{eq:general transitional function}) \emph{does} rely on the value of $b$. Firstly, if the transitional potential is repulsive, namely $b\leqslant0$, then the short-range behaviour of radial wave function is dominated by the $l=0$ mode,
\begin{eqnarray}
u(\rho)\sim u_0(\rho)\sim \rho^{\frac{1}{2}\left(1+\sqrt{1-4b}\right)}\;.
\end{eqnarray}
Secondly, if $b$ satisfies $l_0\left(l_0+1\right)<b\leqslant \frac{1}{4}\left(2l_0+1\right)^2$, where $l_0=0,1,2,\ldots$, then the $l=l_0$ mode dominates at short range and we have 
\begin{eqnarray}
u_l(\rho)\sim u_{l_0}\sim\rho^{\frac{1}{2}\left(1-\sqrt{\left(2l_0+1\right)^2-4b}\right)}\;.
\end{eqnarray}
Finally, if $b$ satisfies $\frac{1}{4}\left(2l_0+1\right)^2<b\leqslant\left(l_0+1\right)\left(l_0+2\right)$, then the short-range behaviour is dominated by all modes no larger than $l_0$
\begin{eqnarray}
u_l(\rho)\sim u_{l^\prime}(\rho)\sim \rho^{\frac{1}{2}\left(1\pm i\sqrt{4b-\left(2l^\prime+1\right)^2}\right)}\;,\quad
l^\prime=0,1,2,\ldots,l_0\;.
\end{eqnarray}
For illustration, we have explicitly listed some values of $b$ in Table~\ref{table} along with the modes that dominate at short ranges and the asymptotic behaviour of the radial wave function as $\rho\to0$.

It is helpful to consider a concrete example. The short-range behaviour of the neutrino force in the $s$-channel case is an attractive transitional potential (cf. Eq.~(\ref{eq:-18})),
\begin{eqnarray}
V_s(r)=-\frac{3\alpha_\nu^2}{32}\frac{1}{m_\chi r^2}=-\frac{3\alpha_\nu^2}{32}\frac{E}{\rho^2}\;,
\end{eqnarray}
thus we have $b=3\alpha_\nu^2/32\approx 0.1\alpha_\nu^2$ in our case. The perturbativity of the theory requires $\alpha_\nu<1$ so $b$ is certainly smaller than $1/4$, which means that the dominant mode is $l=0$. It is then interesting to analyse the asymptotic behaviour of the radial wave function for small $b$. 

When $b$ equals zero, i.e. in the decoupling limit, the two particular solutions of $u_0(\rho)$ are simply $\cos\rho$ and $\sin\rho$, which behave as a constant and as $\rho$ at small distances. For small but nonzero values of $b$, $u_0(\rho)$ is a linear combination of $\rho^{b}$ and $\rho^{1-b}$ at small distances. However, we know from Eq.~(\ref{eq:-14})
that the free solution without the potential behaves as $u_{{\rm free},0}(\rho)\sim\rho$ as $\rho\to0$, so the Sommerfeld enhancement factor is given by
\begin{eqnarray}
S=\lim_{\rho\to0}\left|\frac{u_0(\rho)}{u_{{\rm free},0}(\rho)}\right|^2\sim\rho^{-2b}\;,
\end{eqnarray}
which tends to zero for negative $b$ and to infinity for positive $b$ as $\rho\to0$. This means that there is no Sommerfeld enhancement for a repulsive transitional potential, while for an attractive transitional potential the formalism used to calculate Sommerfeld enhancement in Eq.~(\ref{eq:-9}) no longer holds\footnote{Notice that this is not in contradiction with the results obtained in Sec.~\ref{subsec:s-channel} because as $r$ approaches $m_\chi^{-1}$, the non-relativistic approximation of $\chi$ becomes invalid so one must put a cutoff on the potential when $r\lesssim m_\chi^{-1}$ in the practical computation.}.  

\bibliographystyle{JHEP}
\bibliography{NeutrinoForce}

\providecommand{\href}[2]{#2}\begingroup\raggedright\begin{thebibliography}{10}

\bibitem{Sommerfeld}
A.~Sommerfeld, {\it \"uber die beugung und bremsung der elektronen},  {\em
  Annalen der Physik} {\bf 403} (1931), no.~3 257--330.

\bibitem{Hisano:2003ec}
J.~Hisano, S.~Matsumoto, and M.~M. Nojiri, {\it {Explosive dark matter
  annihilation}},  {\em Phys. Rev. Lett.} {\bf 92} (2004) 031303,
  [\href{http://www.arxiv.org/abs/hep-ph/0307216}{{\tt hep-ph/0307216}}].

\bibitem{Hisano:2004ds}
J.~Hisano, S.~Matsumoto, M.~M. Nojiri, and O.~Saito, {\it {Non-perturbative
  effect on dark matter annihilation and gamma ray signature from galactic
  center}},  {\em Phys. Rev. D} {\bf 71} (2005) 063528,
  [\href{http://www.arxiv.org/abs/hep-ph/0412403}{{\tt hep-ph/0412403}}].

\bibitem{Hisano:2005ec}
J.~Hisano, S.~Matsumoto, O.~Saito, and M.~Senami, {\it {Heavy wino-like
  neutralino dark matter annihilation into antiparticles}},  {\em Phys. Rev. D}
  {\bf 73} (2006) 055004, [\href{http://www.arxiv.org/abs/hep-ph/0511118}{{\tt
  hep-ph/0511118}}].

\bibitem{Hisano:2006nn}
J.~Hisano, S.~Matsumoto, M.~Nagai, O.~Saito, and M.~Senami, {\it
  {Non-perturbative effect on thermal relic abundance of dark matter}},  {\em
  Phys. Lett. B} {\bf 646} (2007) 34--38,
  [\href{http://www.arxiv.org/abs/hep-ph/0610249}{{\tt hep-ph/0610249}}].

\bibitem{Cirelli:2007xd}
M.~Cirelli, A.~Strumia, and M.~Tamburini, {\it {Cosmology and Astrophysics of
  Minimal Dark Matter}},  {\em Nucl. Phys. B} {\bf 787} (2007) 152--175,
  [\href{http://www.arxiv.org/abs/0706.4071}{{\tt 0706.4071}}].

\bibitem{March-Russell:2008lng}
J.~March-Russell, S.~M. West, D.~Cumberbatch, and D.~Hooper, {\it {Heavy Dark
  Matter Through the Higgs Portal}},  {\em JHEP} {\bf 07} (2008) 058,
  [\href{http://www.arxiv.org/abs/0801.3440}{{\tt 0801.3440}}].

\bibitem{PAMELA:2008gwm}
{\bf PAMELA} {\bf Collaboration}, O.~Adriani {\em et~al.}, {\it {An anomalous
  positron abundance in cosmic rays with energies 1.5-100 GeV}},  {\em Nature}
  {\bf 458} (2009) 607--609, [\href{http://www.arxiv.org/abs/0810.4995}{{\tt
  0810.4995}}].

\bibitem{Chang:2008aa}
J.~Chang {\em et~al.}, {\it {An excess of cosmic ray electrons at energies of
  300-800 GeV}},  {\em Nature} {\bf 456} (2008) 362--365.

\bibitem{Fermi-LAT:2009yfs}
{\bf Fermi-LAT} {\bf Collaboration}, A.~A. Abdo {\em et~al.}, {\it {Measurement
  of the Cosmic Ray e+ plus e- spectrum from 20 GeV to 1 TeV with the Fermi
  Large Area Telescope}},  {\em Phys. Rev. Lett.} {\bf 102} (2009) 181101,
  [\href{http://www.arxiv.org/abs/0905.0025}{{\tt 0905.0025}}].

\bibitem{Cirelli:2008jk}
M.~Cirelli and A.~Strumia, {\it {Minimal Dark Matter predictions and the PAMELA
  positron excess}},  {\em PoS} {\bf IDM2008} (2008) 089,
  [\href{http://www.arxiv.org/abs/0808.3867}{{\tt 0808.3867}}].

\bibitem{Cirelli:2008pk}
M.~Cirelli, M.~Kadastik, M.~Raidal, and A.~Strumia, {\it {Model-independent
  implications of the $e+$, $e-$, anti-proton cosmic ray spectra on properties
  of Dark Matter}},  {\em Nucl. Phys. B} {\bf 813} (2009) 1--21,
  [\href{http://www.arxiv.org/abs/0809.2409}{{\tt 0809.2409}}]. [Addendum:
  Nucl.Phys.B 873, 530--533 (2013)].

\bibitem{Arkani-Hamed:2008hhe}
N.~Arkani-Hamed, D.~P. Finkbeiner, T.~R. Slatyer, and N.~Weiner, {\it {A Theory
  of Dark Matter}},  {\em Phys. Rev. D} {\bf 79} (2009) 015014,
  [\href{http://www.arxiv.org/abs/0810.0713}{{\tt 0810.0713}}].

\bibitem{Pospelov:2008jd}
M.~Pospelov and A.~Ritz, {\it {Astrophysical Signatures of Secluded Dark
  Matter}},  {\em Phys. Lett. B} {\bf 671} (2009) 391--397,
  [\href{http://www.arxiv.org/abs/0810.1502}{{\tt 0810.1502}}].

\bibitem{Fox:2008kb}
P.~J. Fox and E.~Poppitz, {\it {Leptophilic Dark Matter}},  {\em Phys. Rev. D}
  {\bf 79} (2009) 083528, [\href{http://www.arxiv.org/abs/0811.0399}{{\tt
  0811.0399}}].

\bibitem{Lattanzi:2008qa}
M.~Lattanzi and J.~I. Silk, {\it {Can the WIMP annihilation boost factor be
  boosted by the Sommerfeld enhancement?}},  {\em Phys. Rev. D} {\bf 79} (2009)
  083523, [\href{http://www.arxiv.org/abs/0812.0360}{{\tt 0812.0360}}].

\bibitem{Pieri:2009zi}
L.~Pieri, M.~Lattanzi, and J.~Silk, {\it {Constraining the Sommerfeld
  enhancement with Cherenkov telescope observations of dwarf galaxies}},  {\em
  Mon. Not. Roy. Astron. Soc.} {\bf 399} (2009) 2033,
  [\href{http://www.arxiv.org/abs/0902.4330}{{\tt 0902.4330}}].

\bibitem{Bovy:2009zs}
J.~Bovy, {\it {Substructure Boosts to Dark Matter Annihilation from Sommerfeld
  Enhancement}},  {\em Phys. Rev. D} {\bf 79} (2009) 083539,
  [\href{http://www.arxiv.org/abs/0903.0413}{{\tt 0903.0413}}].

\bibitem{Yuan:2009bb}
Q.~Yuan, X.-J. Bi, J.~Liu, P.-F. Yin, J.~Zhang, and S.-H. Zhu, {\it {Clumpiness
  enhancement of charged cosmic rays from dark matter annihilation with
  Sommerfeld effect}},  {\em JCAP} {\bf 12} (2009) 011,
  [\href{http://www.arxiv.org/abs/0905.2736}{{\tt 0905.2736}}].

\bibitem{Slatyer:2009vg}
T.~R. Slatyer, {\it {The Sommerfeld enhancement for dark matter with an excited
  state}},  {\em JCAP} {\bf 02} (2010) 028,
  [\href{http://www.arxiv.org/abs/0910.5713}{{\tt 0910.5713}}].

\bibitem{Feng:2009hw}
J.~L. Feng, M.~Kaplinghat, and H.-B. Yu, {\it {Halo Shape and Relic Density
  Exclusions of Sommerfeld-Enhanced Dark Matter Explanations of Cosmic Ray
  Excesses}},  {\em Phys. Rev. Lett.} {\bf 104} (2010) 151301,
  [\href{http://www.arxiv.org/abs/0911.0422}{{\tt 0911.0422}}].

\bibitem{Feng:2010zp}
J.~L. Feng, M.~Kaplinghat, and H.-B. Yu, {\it {Sommerfeld Enhancements for
  Thermal Relic Dark Matter}},  {\em Phys. Rev. D} {\bf 82} (2010) 083525,
  [\href{http://www.arxiv.org/abs/1005.4678}{{\tt 1005.4678}}].

\bibitem{Cholis:2010px}
I.~Cholis and L.~Goodenough, {\it {Consequences of a Dark Disk for the Fermi
  and PAMELA Signals in Theories with a Sommerfeld Enhancement}},  {\em JCAP}
  {\bf 09} (2010) 010, [\href{http://www.arxiv.org/abs/1006.2089}{{\tt
  1006.2089}}].

\bibitem{Zavala:2009mi}
J.~Zavala, M.~Vogelsberger, and S.~D.~M. White, {\it {Relic density and CMB
  constraints on dark matter annihilation with Sommerfeld enhancement}},  {\em
  Phys. Rev. D} {\bf 81} (2010) 083502,
  [\href{http://www.arxiv.org/abs/0910.5221}{{\tt 0910.5221}}].

\bibitem{Hannestad:2010zt}
S.~Hannestad and T.~Tram, {\it {Sommerfeld Enhancement of DM Annihilation:
  Resonance Structure, Freeze-Out and CMB Spectral Bound}},  {\em JCAP} {\bf
  01} (2011) 016, [\href{http://www.arxiv.org/abs/1008.1511}{{\tt 1008.1511}}].

\bibitem{Iminniyaz:2010hy}
H.~Iminniyaz and M.~Kakizaki, {\it {Thermal abundance of non-relativistic
  relics with Sommerfeld enhancement}},  {\em Nucl. Phys. B} {\bf 851} (2011)
  57--65, [\href{http://www.arxiv.org/abs/1008.2905}{{\tt 1008.2905}}].

\bibitem{Iminniyaz:2011pva}
H.~Iminniyaz, X.-L. Chen, X.-J. Bi, and S.~Dulat, {\it {Effects of Kinetic
  Decoupling on Relic Density with Sommerfeld Enhancement}},  {\em Commun.
  Theor. Phys.} {\bf 56} (2011) 967--971.

\bibitem{Hryczuk:2011tq}
A.~Hryczuk, {\it {The Sommerfeld enhancement for scalar particles and
  application to sfermion co-annihilation regions}},  {\em Phys. Lett. B} {\bf
  699} (2011) 271--275, [\href{http://www.arxiv.org/abs/1102.4295}{{\tt
  1102.4295}}].

\bibitem{Iengo:2009xf}
R.~Iengo, {\it {Sommerfeld enhancement for a Yukawa potential}},
  \href{http://www.arxiv.org/abs/0903.0317}{{\tt 0903.0317}}.

\bibitem{Iengo:2009ni}
R.~Iengo, {\it {Sommerfeld enhancement: General results from field theory
  diagrams}},  {\em JHEP} {\bf 05} (2009) 024,
  [\href{http://www.arxiv.org/abs/0902.0688}{{\tt 0902.0688}}].

\bibitem{Cassel:2009wt}
S.~Cassel, {\it {Sommerfeld factor for arbitrary partial wave processes}},
  {\em J. Phys. G} {\bf 37} (2010) 105009,
  [\href{http://www.arxiv.org/abs/0903.5307}{{\tt 0903.5307}}].

\bibitem{Das:2016ced}
A.~Das and B.~Dasgupta, {\it {Selection Rule for Enhanced Dark Matter
  Annihilation}},  {\em Phys. Rev. Lett.} {\bf 118} (2017), no.~25 251101,
  [\href{http://www.arxiv.org/abs/1611.04606}{{\tt 1611.04606}}].

\bibitem{Bedaque:2009ri}
P.~F. Bedaque, M.~I. Buchoff, and R.~K. Mishra, {\it {Sommerfeld enhancement
  from Goldstone pseudo-scalar exchange}},  {\em JHEP} {\bf 11} (2009) 046,
  [\href{http://www.arxiv.org/abs/0907.0235}{{\tt 0907.0235}}].

\bibitem{Chen:2009ch}
C.-H. Chen and C.~S. Kim, {\it {Sommerfeld Enhancement from Unparticle Exchange
  for Dark Matter Annihilation}},  {\em Phys. Lett. B} {\bf 687} (2010)
  232--235, [\href{http://www.arxiv.org/abs/0909.1878}{{\tt 0909.1878}}].

\bibitem{McDonald:2012nc}
K.~L. McDonald, {\it {Sommerfeld Enhancement from Multiple Mediators}},  {\em
  JHEP} {\bf 07} (2012) 145, [\href{http://www.arxiv.org/abs/1203.6341}{{\tt
  1203.6341}}].

\bibitem{Zhang:2013qza}
Z.~Zhang, {\it {Multi-Sommerfeld enhancement in dark sector}},  {\em Phys.
  Lett. B} {\bf 734} (2014) 188--192,
  [\href{http://www.arxiv.org/abs/1307.2206}{{\tt 1307.2206}}]. [Erratum:
  Phys.Lett.B 774, 724--724 (2017)].

\bibitem{Strumia:2008cf}
A.~Strumia, {\it {Sommerfeld corrections to type-II and III leptogenesis}},
  {\em Nucl. Phys. B} {\bf 809} (2009) 308--317,
  [\href{http://www.arxiv.org/abs/0806.1630}{{\tt 0806.1630}}].

\bibitem{Feinberg:1968zz}
G.~Feinberg and J.~Sucher, {\it {Long-Range Forces from Neutrino-Pair
  Exchange}},  {\em Phys. Rev.} {\bf 166} (1968) 1638--1644.

\bibitem{Feinberg:1989ps}
G.~Feinberg, J.~Sucher, and C.~K. Au, {\it The dispersion theory of dispersion
  forces},  {\em Phys. Rept.} {\bf 180} (1989) 83.

\bibitem{Hsu:1992tg}
S.~D.~H. Hsu and P.~Sikivie, {\it Long range forces from two neutrino exchange
  revisited},  {\em Phys. Rev. D} {\bf 49} (1994) 4951--4953,
  [\href{http://www.arxiv.org/abs/hep-ph/9211301}{{\tt hep-ph/9211301}}].

\bibitem{Grifols:1996fk}
J.~A. Grifols, E.~Masso, and R.~Toldra, {\it Majorana neutrinos and long range
  forces},  {\em Phys. Lett. B} {\bf 389} (1996) 563--565,
  [\href{http://www.arxiv.org/abs/hep-ph/9606377}{{\tt hep-ph/9606377}}].

\bibitem{Xu:2021daf}
X.-J. Xu and B.~Yu, {\it {On the short-range behavior of neutrino forces beyond
  the Standard Model: from 1/r$^{5}$ to 1/r$^{4}$, 1/r$^{2}$, and 1/r}},  {\em
  JHEP} {\bf 02} (2022) 008, [\href{http://www.arxiv.org/abs/2112.03060}{{\tt
  2112.03060}}].

\bibitem{Hartle:1970ug}
J.~B. Hartle, {\it Long-range weak forces and cosmology},  {\em Phys. Rev. D}
  {\bf 1} (1970) 394--397.

\bibitem{Horowitz:1993kw}
C.~J. Horowitz and J.~T. Pantaleone, {\it Long range forces from the
  cosmological neutrinos background},  {\em Phys. Lett. B} {\bf 319} (1993)
  186--190, [\href{http://www.arxiv.org/abs/hep-ph/9306222}{{\tt
  hep-ph/9306222}}].

\bibitem{Ferrer:1998ju}
F.~Ferrer, J.~A. Grifols, and M.~Nowakowski, {\it Long range forces induced by
  neutrinos at finite temperature},  {\em Phys. Lett. B} {\bf 446} (1999)
  111--116, [\href{http://www.arxiv.org/abs/hep-ph/9806438}{{\tt
  hep-ph/9806438}}].

\bibitem{Fischbach:1996qf}
E.~Fischbach, {\it Long range forces and neutrino mass},  {\em Annals Phys.}
  {\bf 247} (1996) 213--291,
  [\href{http://www.arxiv.org/abs/hep-ph/9603396}{{\tt hep-ph/9603396}}].

\bibitem{Smirnov:1996vj}
A.~Y. Smirnov and F.~Vissani, {\it Long range neutrino forces and the lower
  bound on neutrino mass},  \href{http://www.arxiv.org/abs/hep-ph/9604443}{{\tt
  hep-ph/9604443}}.

\bibitem{Abada:1996nx}
A.~Abada, M.~B. Gavela, and O.~Pene, {\it To rescue a star},  {\em Phys. Lett.
  B} {\bf 387} (1996) 315--319,
  [\href{http://www.arxiv.org/abs/hep-ph/9605423}{{\tt hep-ph/9605423}}].

\bibitem{Kachelriess:1997cr}
M.~Kachelriess, {\it Neutrino selfenergy and pair creation in neutron stars},
  {\em Phys. Lett. B} {\bf 426} (1998) 89--94,
  [\href{http://www.arxiv.org/abs/hep-ph/9712363}{{\tt hep-ph/9712363}}].

\bibitem{Kiers:1997ty}
K.~Kiers and M.~H.~G. Tytgat, {\it Neutrino ground state in a dense star},
  {\em Phys. Rev. D} {\bf 57} (1998) 5970--5981,
  [\href{http://www.arxiv.org/abs/hep-ph/9712463}{{\tt hep-ph/9712463}}].

\bibitem{Abada:1998ti}
A.~Abada, O.~Pene, and J.~Rodriguez-Quintero, {\it Finite size effects on
  multibody neutrino exchange},  {\em Phys. Rev. D} {\bf 58} (1998) 073001,
  [\href{http://www.arxiv.org/abs/hep-ph/9802393}{{\tt hep-ph/9802393}}].

\bibitem{Arafune:1998ft}
J.~Arafune and Y.~Mimura, {\it Finiteness of multibody neutrino exchange
  potential energy in neutron stars},  {\em Prog. Theor. Phys.} {\bf 100}
  (1998) 1083--1088, [\href{http://www.arxiv.org/abs/hep-ph/9805395}{{\tt
  hep-ph/9805395}}].

\bibitem{Orlofsky:2021mmy}
N.~Orlofsky and Y.~Zhang, {\it {Neutrino as the dark force}},  {\em Phys. Rev.
  D} {\bf 104} (2021), no.~7 075010,
  [\href{http://www.arxiv.org/abs/2106.08339}{{\tt 2106.08339}}].

\bibitem{Wilkinson:2014ksa}
R.~J. Wilkinson, C.~Boehm, and J.~Lesgourgues, {\it {Constraining Dark
  Matter-Neutrino Interactions using the CMB and Large-Scale Structure}},  {\em
  JCAP} {\bf 05} (2014) 011, [\href{http://www.arxiv.org/abs/1401.7597}{{\tt
  1401.7597}}].

\bibitem{Bertoni:2014mva}
B.~Bertoni, S.~Ipek, D.~McKeen, and A.~E. Nelson, {\it {Constraints and
  consequences of reducing small scale structure via large dark matter-neutrino
  interactions}},  {\em JHEP} {\bf 04} (2015) 170,
  [\href{http://www.arxiv.org/abs/1412.3113}{{\tt 1412.3113}}].

\bibitem{Olivares-DelCampo:2017feq}
A.~Olivares-Del~Campo, C.~B\oe{}hm, S.~Palomares-Ruiz, and S.~Pascoli, {\it
  {Dark matter-neutrino interactions through the lens of their cosmological
  implications}},  {\em Phys. Rev. D} {\bf 97} (2018), no.~7 075039,
  [\href{http://www.arxiv.org/abs/1711.05283}{{\tt 1711.05283}}].

\bibitem{Berlin:2017ftj}
A.~Berlin and N.~Blinov, {\it {Thermal Dark Matter Below an MeV}},  {\em Phys.
  Rev. Lett.} {\bf 120} (2018), no.~2 021801,
  [\href{http://www.arxiv.org/abs/1706.07046}{{\tt 1706.07046}}].

\bibitem{Stadler:2019dii}
J.~Stadler, C.~B\oe{}hm, and O.~Mena, {\it {Comprehensive Study of
  Neutrino-Dark Matter Mixed Damping}},  {\em JCAP} {\bf 08} (2019) 014,
  [\href{http://www.arxiv.org/abs/1903.00540}{{\tt 1903.00540}}].

\bibitem{Hufnagel:2021pso}
M.~Hufnagel and X.-J. Xu, {\it {Dark matter produced from neutrinos}},  {\em
  JCAP} {\bf 01} (2022), no.~01 043,
  [\href{http://www.arxiv.org/abs/2110.09883}{{\tt 2110.09883}}].

\bibitem{Hooper:2021rjc}
D.~C. Hooper and M.~Lucca, {\it {Hints of dark matter-neutrino interactions in
  Lyman-$\alpha$ data}},  \href{http://www.arxiv.org/abs/2110.04024}{{\tt
  2110.04024}}.

\bibitem{Blum:2016nrz}
K.~Blum, R.~Sato, and T.~R. Slatyer, {\it {Self-consistent Calculation of the
  Sommerfeld Enhancement}},  {\em JCAP} {\bf 06} (2016) 021,
  [\href{http://www.arxiv.org/abs/1603.01383}{{\tt 1603.01383}}].

\bibitem{Frank:1971xx}
W.~Frank, D.~J. Land, and R.~M. Spector, {\it {Singular potentials}},  {\em
  Rev. Mod. Phys.} {\bf 43} (1971) 36--98.

\bibitem{Landau}
L. D, Landau and E. M. Lifshitz, {\it Mechanics} (Pergamon, London, 1958); {\it
  Quantum Mechanics} (Pergamon, London, 1960).

\bibitem{Fields:2019pfx}
B.~D. Fields, K.~A. Olive, T.-H. Yeh, and C.~Young, {\it {Big-Bang
  Nucleosynthesis after Planck}},  {\em JCAP} {\bf 03} (2020) 010,
  [\href{http://www.arxiv.org/abs/1912.01132}{{\tt 1912.01132}}]. [Erratum:
  JCAP 11, E02 (2020)].

\bibitem{CMS-PAS-HIG-16-036}
{\bf CMS Collaboration} {\bf Collaboration}, {\it {A search for doubly-charged
  Higgs boson production in three and four lepton final states at
  $\sqrt{s}=13~\mathrm{TeV}$}},  tech. rep., CERN, Geneva, 2017.
\newblock CMS-PAS-HIG-16-036.

\bibitem{Abbiendi:2013hk}
{\bf ALEPH, DELPHI, L3, OPAL, LEP} {\bf Collaboration}, G.~Abbiendi {\em
  et~al.}, {\it {Search for Charged Higgs bosons: Combined Results Using LEP
  Data}},  {\em Eur. Phys. J. C} {\bf 73} (2013) 2463,
  [\href{http://www.arxiv.org/abs/1301.6065}{{\tt 1301.6065}}].

\bibitem{Cirelli:2005uq}
M.~Cirelli, N.~Fornengo, and A.~Strumia, {\it {Minimal dark matter}},  {\em
  Nucl. Phys. B} {\bf 753} (2006) 178--194,
  [\href{http://www.arxiv.org/abs/hep-ph/0512090}{{\tt hep-ph/0512090}}].

\bibitem{ParticleDataGroup:2020ssz}
{\bf Particle Data Group} {\bf Collaboration}, P.~A. Zyla {\em et~al.}, {\it
  {Review of Particle Physics}},  {\em PTEP} {\bf 2020} (2020), no.~8 083C01.

\bibitem{Chikashige:1980ui}
Y.~Chikashige, R.~N. Mohapatra, and R.~D. Peccei, {\it {Are There Real
  Goldstone Bosons Associated with Broken Lepton Number?}},  {\em Phys. Lett.}
  {\bf 98B} (1981) 265--268.

\bibitem{Xu:2020qek}
X.-J. Xu, {\it {The $\nu_{R}$-philic scalar: its loop-induced interactions and
  Yukawa forces in LIGO observations}},  {\em JHEP} {\bf 09} (2020) 105,
  [\href{http://www.arxiv.org/abs/2007.01893}{{\tt 2007.01893}}].

\bibitem{Fernandez-Martinez:2016lgt}
E.~Fernandez-Martinez, J.~Hernandez-Garcia, and J.~Lopez-Pavon, {\it {Global
  constraints on heavy neutrino mixing}},  {\em JHEP} {\bf 08} (2016) 033,
  [\href{http://www.arxiv.org/abs/1605.08774}{{\tt 1605.08774}}].

\bibitem{Coy:2021wfs}
R.~Coy and X.-J. Xu, {\it {Probing the muon g \ensuremath{-} 2 with future beam
  dump experiments}},  {\em JHEP} {\bf 10} (2021) 189,
  [\href{http://www.arxiv.org/abs/2108.05147}{{\tt 2108.05147}}].

\bibitem{Arguelles:2019ouk}
C.~A. Arg\"uelles, A.~Diaz, A.~Kheirandish, A.~Olivares-Del-Campo, I.~Safa, and
  A.~C. Vincent, {\it {Dark matter annihilation to neutrinos}},  {\em Rev. Mod.
  Phys.} {\bf 93} (2021), no.~3 035007,
  [\href{http://www.arxiv.org/abs/1912.09486}{{\tt 1912.09486}}].

\bibitem{Tulin:2017ara}
S.~Tulin and H.-B. Yu, {\it {Dark Matter Self-interactions and Small Scale
  Structure}},  {\em Phys. Rept.} {\bf 730} (2018) 1--57,
  [\href{http://www.arxiv.org/abs/1705.02358}{{\tt 1705.02358}}].

\bibitem{Case:1950an}
K.~M. Case, {\it {Singular potentials}},  {\em Phys. Rev.} {\bf 80} (1950)
  797--806.

\bibitem{giudice1965singular}
E.~D. Giudice and E.~Galzenati, {\it On singular potential scattering.-i},
  {\em Il Nuovo Cimento (1955-1965)} {\bf 38} (1965), no.~1 443--458.

\bibitem{giudice1965singular2}
E.~D. Giudice and E.~Galzenati, {\it On singular potential scattering.-ii},
  {\em Il Nuovo Cimento A (1965-1970)} {\bf 40} (1965), no.~3 739--747.

\bibitem{Teschlordinary2012}
G.~Teschl, {\em Ordinary differential equations and dynamical systems},
  vol.~140.
\newblock American Mathematical Soc., 2012.

\end{thebibliography}\endgroup

\end{document}